\documentclass[a4paper,10pt]{article}

\usepackage{multirow}	
\usepackage[latin1]{inputenc}
\usepackage[T1]{fontenc}
\usepackage[british]{babel}
\usepackage{graphicx}
\usepackage[dvips]{geometry}
\usepackage{float}
\usepackage{array}
\usepackage{amsmath}
\usepackage{amsfonts}

\usepackage{psfrag}
\usepackage{graphicx}
\usepackage{amssymb}
\usepackage{amsmath}
\usepackage{rotating}
\usepackage{pstricks,pst-fill,pst-text,pst-node,pstricks-add}
\usepackage{lmodern}

\usepackage{color}

\makeatletter
\@addtoreset{equation}{section}
\makeatother

\newcommand{\Ket}[1]{\left|#1  \right>}

\newcommand{\Braket}[1]{\left<#1  \right>}

\newcommand{\proja}[0]{
 \circ \rightarrow \bullet
}

\newcommand{\projb}[0]{
 \bullet
}

\usepackage{rotating}
\usepackage{multirow}

\begin{document}

\title{Indecomposability parameters in chiral Logarithmic Conformal Field Theory}

\author{Romain Vasseur$^{1,2}$, Jesper Lykke Jacobsen$^{2,3}$ and Hubert Saleur$^{1,4}$ \\
[2.0mm]
  ${}^1$Institut de Physique Th\'eorique, CEA Saclay,
  91191 Gif Sur Yvette, France \\
  ${}^2$LPTENS, 24 rue Lhomond, 75231 Paris, France \\
  ${}^3$ Universit\'e Pierre et Marie Curie, 4 place Jussieu, 75252 Paris, France \\
  ${}^4$Department of Physics,
  University of Southern California, Los Angeles, CA 90089-0484}


\maketitle

\begin{abstract}

Work of the last few years has shown that the key algebraic features of Logarithmic Conformal Field Theories (LCFTs)
are already present in some finite lattice systems (such as the XXZ spin-$1/2$ chain) before the continuum limit is taken. This has provided a very convenient 
way to analyze the structure of indecomposable Virasoro modules and to obtain fusion rules for a variety of models such as (boundary) percolation etc. 

LCFTs allow for additional quantum numbers describing the fine structure of the indecomposable modules, and generalizing the `$b$-number' introduced initially 
by Gurarie for the $c=0$ case. The determination of these indecomposability parameters (or logarithmic couplings) has given rise to a lot of algebraic work, but 
their physical meaning has remained somewhat elusive. In a recent paper, a way to measure $b$ for boundary percolation and polymers was proposed. We generalize
 this work here by devising a general strategy to compute matrix elements of Virasoro generators from  the numerical analysis of lattice models and their continuum limit. 
The method is applied to XXZ spin-1/2 and spin-1 chains with
open (free) boundary conditions. They are related to gl($n+m|m$) and 
osp($n+2m|2m$)-invariant superspin chains and to non-linear sigma models with 
supercoset target spaces. These models can also be formulated in terms of dense
and dilute loop gas. 


We check the method in many cases where the results were already known analytically.
Furthermore, we also confront our findings with a construction generalizing Gurarie's, 
where logarithms emerge naturally in operator product expansions to compensate for  apparently divergent terms. 
This argument actually allows us to compute 
indecomposability parameters in any logarithmic theory. A central result 
of our study is the construction of a Kac table for the indecomposability parameters 
of the logarithmic minimal models $\mathcal{LM}(1,p)$ and $\mathcal{LM}(p,p+1)$.

\end{abstract}



\section*{Introduction}

Logarithmic Conformal Field Theories (LCFTs) are characterized by the fact that the dilatation operator 
$L_0$ is non-diagonalizable and has a Jordan cell structure. This property leads to the appearance of logarithms
in correlation functions and to indecomposable operator product expansions. Such theories have emerged recently
in many physical situations such as geometrical problems, disordered free fermions models in 2+1 dimensions 
or the AdS/CFT correspondence in string theory.

Geometrical applications include self-avoiding random walks (polymers) or percolation (see {\it e.g.} \cite{SaleurPoly,MathieuRidout}). 
Although it would be fair to say to these problems are well understood, a fully consistent field theory
describing geometrical observables ({\it e.g.}, four point correlation functions in the bulk) is still missing. Another very exciting question, 
with considerable applications
to condensed matter, concerns the critical point in non-interacting  disordered fermion models in 2+1 dimensions, which 
are believed to be described by $c=0$ LCFTs. Physical realizations of such systems are given by transition between 
plateaux in the  Integer/Spin Quantum Hall Effects. The appearance of logarithmic correlators in 
such theories cannot be avoided \cite{Cardylog} and a deep understanding of LCFTs is necessary in order to describe 
their low-energy physics. While much has been done about the Spin Quantum Hall Effect, as it can be related to the 
classical percolation problem \cite{SQHE}, very little is known about the field theory describing the transition in 
the Integer Quantum Hall Effect (see \cite{Zirnbauer} for a review). Logarithms in CFTs also arise naturally in the context 
 of non-linear sigma models with super target space. These quantum field theories play a major role in the AdS/CFT duality.
For example, the PSL($2|2$) sigma model is related \cite{strings} to strings living in $\mathrm{AdS}_3 \times S^3$.

The key feature of logarithmic CFTs, as opposed to simpler non-unitary theories (such as the Yang-Lee singularity, or other non-unitary 
minimal models), is indecomposability.
This property was probably observed first by Rozansky and Saleur \cite{RozanskySaleur} who studied a WZW model with U($1|1$) supergroup
symmetry and vanishing central charge. They related the non-simplicity of U($1|1$) to the possibility 
of non-diagonalizability of $L_0$ and logarithmic dependence in four-point correlation functions.
The study of logarithmic $c=0$ theories in a more systematic fashion then begun with a serie of papers
by Gurarie and Ludwig. Gurarie \cite{Gurarie1} first noticed that logarithmic operators were necessary in order to construct
a consistent field theory at $c=0$. A similar observation was made by Cardy using a replica approach \cite{Cardylog}. 
Gurarie \cite{Gurarie2} and Gurarie and Ludwig \cite{GurarieLudwig} then related the existence of a logarithmic
partner $t(z)$ for the stress energy tensor $T(z)$ to ill-defined terms in operator product expansions. They introduced one of the
first\footnote{Several other indecomposability parameters were computed by Kausch and Gaberdiel a few
years earlier \cite{KauschGaberdiel}, in a different context.} indecomposability parameters, usually denoted $b$, which at the time 
was interpreted as a new `anomaly' that 
would play the role of a central charge when $c=0$. Using some heuristic arguments, they were also able to predict two possible
values, $b=-\frac{5}{8}$ and $b=\frac{5}{6}$, which would distinguish between two fundamentally different LCFTs.

Instead of a single parameter $b$, it is now well accepted that a LCFT is characterized rather by a complex
structure of indecomposable Virasoro modules, with a infinite number of indecomposability parameters needed to describe the whole pattern. 
Two lines of thought have been considered. The first one is to deal directly with abstract indecomposable Virasoro modules
to try to understand and classify their structures \cite{Rohsiepe,KauschGaberdiel, MathieuRidout,MathieuRidout1,KytolaRidout}.
Progress has been steady in this direction, and the module involving $T$ and its partner $t$ at $c=0$ now appears
as a particular case. These algebraic studies have led to new predictions for indecomposability parameters for various 
logarithmic pairs. The second idea is somehow more concrete, and consists in studying directly lattice models which
can be thought of as lattice regularizations of LCFTs. For specific values of the parameters, the lattice Hamiltonians 
are non-diagonalizable and have a Jordan cell structure that mimics that of the continuum theory. This was mainly 
done by Read and Saleur \cite{RS3}, and independently by Pearce, Rasmussen and Zuber \cite{PRZ,RP1,RP2}. This 
approach essentially relies on the `similarity' between the indecomposable modules of the Temperley-Lieb algebra 
and those of the Virasoro algebra. Both approaches yield a consistent picture of boundary (chiral) LCFTs, with 
common algebraic structure and fusion rules deduced from both methods. As for the similarity, it is now better understood in 
terms of common quantum group structures \cite{Azat}.

While the global structure of the  Virasoro modules in the case of boundary LCFTs is slowly getting under control, 
many questions remain about the existence and values of the parameters arising from indecomposability. It is, 
moreover, tempting to think of these parameters as extensions 
of the structure constants of the operator product algebra of the theory, 
and thus to wonder what they encode physically, and whether it is possible to access them, 
experimentally or at least numerically, in the context of lattice simulations. 
 Recently, Dubail, Jacobsen and Saleur \cite{DJS} suggested a concrete 
method to measure {\sl one}  indecomposability parameter: the  $b$-number of Gurarie in the case of $c=0$ theories.
Their method involved a $c=0$ specific approach, the so called  `trousers trick', and led to the measurement of $b$ for percolation and polymers. 
The observed values---$b=-\frac{5}{8}$  and  $b=\frac{5}{6}$ respectively, were found  in agreement 
with the predictions of Mathieu and Ridout \cite{MathieuRidout}.

The method of \cite{DJS} is geometrically very appealing, but unfortunately does not extend to many other LCFTs, nor 
does it allow the study of indecomposable modules occurring at larger values of the conformal weight. It also does 
not seem to be generalizable to the bulk (non-chiral) case. On the other hand, having at one's disposal a `probe' 
to investigate the detailed structure of modules in LCFTs seems rather essential to make progress in this, so far, 
very abstract topic. We have thus reconsidered the problem using a different route, which involves the identification 
of a lattice stress energy tensor. The original idea for doing so goes back at least to a paper by Koo and Saleur \cite{KooSaleur}, 
who themselves generalized the pioneering work of Kadanoff and Ceva \cite{KadanoffCeva}. The upshot of the proposal in \cite{KooSaleur} 
was that the Temperley-Lieb algebra does not only reproduce {\sl in finite size} indecomposable modules that mimic exactly several 
indecomposables of the Virasoro algebra: on top of this, there exist (infinitely many) linear combinations of words in the Temperley-Lieb 
algebra whose action in these modules reproduces, once properly interpreted, the action of all the Virasoro generators 
{\sl in the thermodynamic limit}. It thus should be possible, in principle, to reconstruct from the lattice all states of the boundary 
LCFT, and to measure all the matrix elements of all the Virasoro generators, hence to determine `experimentally' all the information 
about the Virasoro modules in the model. 

Of course, this program is very difficult to implement in practice. Thanks in part to the progress accomplished in \cite{RS3,DJS}, 
it is however not impossible, and this is what we do in this paper. 

Once the general strategy is under control, it turns out  that we can study many more cases than those considered so far in papers 
on indecomposable Virasoro modules \cite{KauschGaberdiel, MathieuRidout1,KytolaRidout}. 
These authors used mainly two different methods to compute indecomposability parameters. One is based on
the so-called Nahm-Gaberdiel-Kausch algorithm to compute fusion products between indecomposable Virasoro modules \cite{KauschGaberdiel}; 
while the other, used in Refs. \cite{MathieuRidout1,KytolaRidout}, consider different quotients of `glueing' of two Verma modules, 
incomposability parameters are then obtained as a solution of singular-vector equations. Note also that they can be computed using 
SLE considerations \cite{Kytola}, or even directly by solving the differential equations satisfied by several four-point 
correlation functions. The latter method was one of the tools used by Gurarie and Ludwig \cite{GurarieLudwig2} to study the 
case of the stress energy tensor at $c=0$. While it should be possible to generalize the methods of these standard references 
to analyze the Jordan cells and calculate the indecomposability parameters in our cases, we find it more convenient 
here to use a `short cut' analysis based on a generalization of the original arguments of Gurarie for the $c=0$ case. This is 
discussed, as a preamble,  in the first section. The general framework is then discussed in section 2, where we introduce the lattice models, and 
 use  and generalize the algebraic arguments of  Ref. \cite{RS3} to deduce  general indecomposable structures. The next two sections are then
dedicated to numerical measurements of indecomposability parameters. All our results are consistent with  the 
predictions  in \cite{KauschGaberdiel, MathieuRidout1,KytolaRidout}, but we  obtain many results beyond these references---those 
for the first few logarithmic minimal models $\mathcal{LM}(1,p)$ and $\mathcal{LM}(p,p+1)$ are summarized in table \ref{general_b_figure}. 
A few conclusions are finally gathered in the last section.


\section{Preamble: indecomposability parameters and $\epsilon \rightarrow 0$ argument}

In this section, we define in simple terms the indecomposability parameter associated 
with a general pair of logarithmic operators. We then extend the $c \rightarrow 0$ 
argument of Gurarie to other logarithmic CFTs and we show that this allows us to predict 
the value of the indecomposability parameters of a given theory.

\subsection{Jordan cells and indecomposability parameters}

It is now well known that the appearance of logarithms in correlation function is related 
to the non-diagonalizability of the $L_0$ operator \cite{Gurarie1}. This Jordan cell 
structure of the Hamiltonian ($L_0$) is itself deeply related to the non-semisimplicity of 
the underlying symmetry algebra of the theory. We will come back to these algebraic 
considerations later and choose here to focus on the Jordan cell structure. Consider 
a pair of logarithmic operators ($\phi(z)$,$\psi(z)$) with conformal weight $h$ that 
are mixed by $L_0$ into a Jordan cell. In the basis ($\phi$,$\psi$), the generator of 
the scale transformation reads
\begin{equation}
L_0 = 
\left( \begin{array}{cc}
h & 1 \\
0 & h \end{array} \right).
\end{equation}
Invariance under global conformal transformations then fixes \cite{Gurarie1} the form 
of the correlation functions
\begin{subequations}
\begin{eqnarray}
\left\langle \phi(z) \phi(0)\right\rangle &=& 0 \\
\left\langle \phi(z) \psi(0)\right\rangle &=& \frac{\beta}{z^{2h}} \\
\left\langle \psi(z) \psi(0)\right\rangle &=& \frac{\theta -2 \beta \log z} {z^{2h}} ,
\end{eqnarray}
\end{subequations}
where $\theta$ and $\beta$ are two parameters. While the constant $\theta$ is arbitrary 
and can be canceled by a choice $\psi \rightarrow \psi - \frac{\theta}{2 \beta} \phi $, the parameter 
$\beta$ is a fundamental number that characterizes the structure of the Jordan cell. It is 
also important to remark that $\phi(z)$ must be a null-field by conformal invariance, that 
is to say, introducing the usual Virasoro bilinear form, $\Braket{\phi|\phi} =0$. 
Actually, we know more about the algebraic structure hidden behind the non-diagonalizability 
of $L_0$. As we will see in details in the following, the fields $\psi(z)$ and $\phi(z)$ always 
appear at the top and the bottom of a larger structure called a staggered Virasoro module \cite{MathieuRidout,Rohsiepe,KytolaRidout}, 
also called projective\footnote{Note that this nomenclature is somewhat dangerous as strictly speaking, these modules 
may not be projective over the Virasoro algebra in the mathematical sense. However, they can be seen as 
`scaling limit' of projective modules over the Temperley-Lieb algebra that arises in lattice models ({\it cf.} next section),
it is thus very tempting to call them projective anyway.} module in Ref \cite{RS3}. The fields 
in these modules are organized in a diamond-shaped structure that we note
\begin{equation}
\mathcal{P}=
 \begin{array}{ccccc}
      &&\hskip-.7cm \psi &&\\
      &\hskip-.2cm\swarrow&\searrow&\\
      \xi &&&\hskip-.3cm \rho \\
      &\hskip-.2cm\searrow&\swarrow&\\
      &&\hskip-.7cm \phi &&
\end{array}
.
\label{eq_general_proj}
\end{equation}
The arrows represent the action of Virasoro generators so that the whole structure can be 
induced by action of the $L_n$'s on the field $\psi(z)$ while $\phi(z)$ belongs to an invariant 
submodule. The whole module is reducible but indecomposable under the action of the Virasoro 
algebra.  Remark that $L_0$ maps $\psi$ onto $\phi$ due to the Jordan cell action. The conformal 
weights of the different fields in~\eqref{eq_general_proj} satisfy 
$h_{\xi} \leq h=h_{\phi}=h_{\psi} \leq h_{\rho}$. The field $\phi(z)$ is a null descendant of $\xi$
\begin{equation}
\displaystyle \phi(z) = A \xi(z), \ \ A = L_{-n} + \alpha_{(1)} L_{-n+1}L_{-1}+ \dots + \alpha_{(P(n)-1)} L_{-1}^n,
\label{eq_Aconv}
\end{equation}
where $n = h-h_{\xi}$ and $P(n)$ is the number of partitions of the integer $n$. The $\alpha_{(i)}$ 
coefficients are uniquely fixed by the null-vector condition $L_{+1} \phi = L_{+2} \phi =0 $.

If two fields $\psi(z)$ and $\phi(z)$ satisfy all these relations, we say that $\psi(z)$ is the 
logarithmic partner of the null-field $\phi(z)$, with {\it indecomposability parameter} (also 
called logarithmic coupling) $\beta$. Using the Virasoro scalar product, we see that 
\begin{equation}
\displaystyle \beta = \Braket{\psi | \phi}.
\end{equation}
where we normalized $\xi(z)$ such that $\Braket{\xi | \xi} = 1$. Note that it is also possible 
to define $\beta$ through the equation
\begin{equation}
\displaystyle A^\dag \psi (z) = \beta \xi (z).
\end{equation}
It is important to notice at this point that the choice that we adopted for the normalization of the operator $A$
is crucial for the value of $\beta$. Different choices have been used in the literature, and some of them may
yield simpler expressions for $\beta$. Unless otherwise indicated, we always use the convention given by eq.~\eqref{eq_Aconv},
which is consistent with the normalization of the stress energy tensor in $c=0$ theories: $T=L_{-2} I$ so $A=L_{-2}$ 
(see next paragraph). 

Finally, let us remark that the field $\rho(z)$ never enters the computations of $\beta$ so that
it can be ignored as far as indecomposability parameters are concerned.

\subsection{$c \rightarrow 0$ catastrophe and the stress energy tensor}

\label{subsec_catastrophe}
We now  show that 
the $\beta$'s are fixed by a very simple argument relying on operator product 
expansions (OPEs). The basic idea was developed in \cite{Gurarie1,GurarieLudwig}, and a similar discussion may be found in \cite{KoganNichols,Cardylog2}. Consider a general CFT with the following parametrization 
of the central charge and of the Kac formula
\begin{subequations}
\begin{eqnarray}
\displaystyle c &=& 1 - \frac{6}{x (x+1)}, \\
\displaystyle h_{r,s} &=& \frac{ \left[(x+1)r - xs \right]^2 - 1}{4 x (x+1)}.
\end{eqnarray}
\end{subequations}
Conformal invariance fixes the OPE of an operator $\Phi_h(z)$ with itself to be of the form
\begin{equation}
\displaystyle \Phi_h(z) \Phi_h(0) \sim \frac{a_\Phi}{z^{2h}} \left[1 + \frac{2 h}{c} z^2 T(0) +\dots \right],
\end{equation}
where $T(z) = L_{-2} I$ is the stress energy tensor of the theory. This expression is 
clearly ill-defined as $c \rightarrow 0$. In a nutshell, the idea of Gurarie was to 
introduce another field $\Phi_{1,5}$ for percolation ({\it resp.} $\Phi_{3,1}$ for 
dilute polymers) with conformal weight $h_t \equiv h_{1,5}=2$ ({\it resp.} 
$h_t \equiv h_{3,1}=2$) at $c=0$ to cancel this divergence\footnote{Actually, the 
original guess of Gurarie and Ludwig was $\Phi_{1,5}$ for dilute polymers and $\Phi_{3,1}$ 
for percolation. It shall become clear in the next sections that this result was correct 
up to a switch, as already remarked in Refs. \cite{MathieuRidout, DJS} .}. Let us focus on the percolation case. When $c$ is slightly different 
from $0$ ($x = 2 + \epsilon$), we can normalize\footnote{For the sake of clarity, we absorb the coefficient $C^{\Phi_{1,5}}_{\Phi,\Phi}/a_{\Phi}$ into
the normalization of $\Phi_{1,5}$ as this will play no role in the following. The coefficients $C^{k}_{i,j}$ are the usual structure constants that appear in OPEs; note also that $a_{\Phi} = C^{\Phi_{1,1}}_{\Phi,\Phi}$.} the field $\Phi_{1,5}$ such that the OPE reads
\begin{equation}
\displaystyle \Phi_h(z) \Phi_h(0) \sim \frac{a_\Phi}{z^{2h}} \left[1 + \frac{2 h}{c} z^2 T(0) + z^{h_t}  \Phi_{1,5} (0)+ \dots \right].
\end{equation}
We then define a new field $t(z)$ as
\begin{equation}\label{eq_newfieldt}
\displaystyle  \Phi_{1,5} (z)  = \frac{2h \Braket{T | T}}{c \beta(\epsilon)} t(z) - \frac{2h}{c} T(z),
\end{equation}
where $\beta(\epsilon) = - \frac{ \Braket{T | T}}{h_{t}-2}$ and $\Braket{T | T}=\frac{c}{2}$. The OPE then involves quantities that are perfectly well defined as $c\rightarrow 0$
\begin{equation}
\displaystyle \Phi_h(z) \Phi_h(0) \sim \frac{a_\Phi}{z^{2h}} \left[1 + \frac{ h}{\beta} z^2( T(0) \log z + t(0)) + \dots \right],
\end{equation}
with $\beta=\lim_{\epsilon \rightarrow 0} \beta (\epsilon)$. 
It is important to realize that the new field $t(z)$ is perfectly well defined as  $c\rightarrow 0$, while the $L_0$ eigenvector $\Phi_{1,5}$
is not. In particular, one can then check 
(see {\it e.g.} \cite{KoganNichols}) that the fields $T(z)$ and $t(z)$ then satisfy 
the standard equations for logarithmic operators
\begin{subequations}
\begin{eqnarray}
\left\langle T(z) T(0)\right\rangle &=& 0 \\
\left\langle T(z) t(0)\right\rangle &=& \frac{\beta}{z^4} \\
\left\langle t(z) t(0)\right\rangle &=& \frac{\theta -2 \beta \log z}{z^4} ,
\end{eqnarray}
\end{subequations}
with $\theta$ a constant. A straightforward calculation using eq.~\eqref{eq_newfieldt} also shows that $L_{0} t = 2 t + T$ as
expected.
Note that it is quite general that one of the eigenvectors (here $\Phi_{1,5}$) of an operator (here $L_0$) 
diverges as one tunes a parameter (here $\epsilon$) to approach an indecomposable point. One can then construct a new 
Jordan vector (here $t$) by canceling the diverging part in the ill-defined eigenvector by taking an appropriate combination with the
eigenvector that has the same eigenvalue at the indecomposable point (here $T$).

 $T(z)$ and its logarithmic partner $t(z)$ are a special 
case of the general structure described in the previous paragraph. In particular, 
they are organized in a diamond structure like~\eqref{eq_general_proj} with 
$\xi = I$, $\phi = T$, $\psi = t$, $A = L_{-2}$ and $L_{2}t = \beta I$.

Gurarie and Ludwig then inferred the value of $\beta$ using 
algebraic arguments along with some heuristic hypotheses \cite{GurarieLudwig}. 
At this point, it is important to notice that it is also possible to compute $\beta$ using the simple limit process.
This was already noticed by Gurarie and Ludwig in the context of the replica approach \cite{GurarieLudwig2}.
Using the parametrization $x = 2 + \epsilon$, we find
\begin{subequations}
\begin{eqnarray}
\beta_{\mathrm{percolation}} &=& - \lim_{\epsilon \rightarrow 0} \frac{c/2}{h_{1,5}-2} =  -\frac{5}{8}, \\
\beta_{\mathrm{polymers}} &=& - \lim_{\epsilon \rightarrow 0} \frac{c/2}{h_{3,1}-2} =  \frac{5}{6},
\end{eqnarray}
\end{subequations}
which indeed are the expected values \cite{MathieuRidout, DJS}. In this sense, 
it is possible to understand the structure of a logarithmic CFT as a limit of 
non-logarithmic CFTs. We suspect that the values of indecomposability parameters 
can be inferred in a similar fashion for any LCFT. To see this, we turn to a 
slightly more complicated example.

\subsection{Generalization to other LCFTs and general formula for $\beta$}

Let us consider a generalization of this argument to a more complicated case. 
We focus in this paragraph on the theory of symplectic fermions \cite{Kausch1,Kausch2} 
that describes the scaling limit of dense polymers on the lattice. This theory has 
$c=-2$ $(x=1)$, and is particularly simple as there is no interaction. The Jordan 
cell structure can be understood in terms of free fermions, and the indecomposability 
parameters can be, in principle, computed using this free fermion representation. This 
theory was also considered in order to test the Nahm-Gaberdiel-Kausch algorithm 
to compute fusion products between indecomposable Virasoro modules \cite{KauschGaberdiel}.

We choose here to focus on a concrete example: this theory is known to have a Jordan 
cell at level 3 with equations
\begin{align}
L_0 \phi & = 3 \phi \\ \notag
L_0 \psi & = 3 \psi + \phi \\ \notag
\phi & = A \xi \\ \notag
A^\dag \psi & = \beta \xi \\ \notag
A &= L_{-1}^2 - 2 L_{-2},
\end{align}
where the parameter $\beta=\Braket{\phi|\psi}=-18$ is the logarithmic coupling 
associated with this Jordan cell \cite{KauschGaberdiel}. Note that we chose a 
different normalization convention for the operator $A$ than eq.~\eqref{eq_Aconv} in order 
to match \cite{KauschGaberdiel}.
When we think of this theory as the continuum limit of a XX spin chain with 
quantum $U_{q=i}(\mathfrak{sl}_2)$ symmetry with an even number of sites \cite{RS3}, the 
conformal dimensions that appear in the spectrum are $h_{1,1+2j}$ with 
$j \in \mathbb{N}$. We shall come back to the precise nature of the scaling 
limit of such lattice models in the next section. All we need for what follows 
is to identify the conformal weights of the fields $\xi$ and $\psi$ in the 
spectrum. We find $h_{\xi} = h_{1,5} = 1$ and $h_{\psi} = h_{1,7} = 3$. It is 
important to identify the operators in spectrum as we are interested in small 
perturbations around $c=-2$. Let us now consider a conformal field theory slightly 
deformed from $c=-2$, with $x=1+\epsilon$. Within this generic (non-logarithmic) 
CFT, the OPE of a generic operator $\Phi_h$ with itself reads
\begin{equation}
\displaystyle \Phi_h(z) \Phi_h(0) \sim \frac{a_\Phi}{z^{2h-h_\xi}} \left[\xi(0) +\frac{1}{2} z \partial \xi(0) + \alpha^{(-2)} z^2 L_{-2} \xi(0) + \alpha^{(-1,-1)} z^2 L_{-1}^2 \xi(0) +\dots \right],
\end{equation}
where the $\alpha$ coefficients are fixed by conformal invariance and are 
diverging as $\epsilon \rightarrow 0$: $\alpha^{(-2)} = \frac{4h}{27 \epsilon}+\frac{1+2h}{27} + \mathcal{O}(\epsilon)$ and $\alpha^{(-1,-1)} = -\frac{2h}{27 \epsilon}+\frac{4 + h}{27} + \mathcal{O}(\epsilon)$. Note that we showed only the channel 
of $\xi$ (with dimension $h_\xi = h_{1,5} = 1 + \frac{3}{2} \epsilon + \mathcal{O}(\epsilon^2)$)
on the right-hand side. Let us 
now introduce the field $\phi = (L_{-1}^2 - 2 L_{-2}) \xi$
\begin{equation}
\displaystyle \Phi_h(z) \Phi_h(0) \sim \frac{a_\Phi}{z^{2h-h_\xi}} \left[\xi(0) +\frac{1}{2} z \partial \xi(0) + \alpha^{(-1,-1)} z^2 \phi(0) +   (2 \alpha^{(-1,-1)}+ \alpha^{(-2)})z^2 L_{-2} \xi(0) +\dots \right].
\end{equation}
Remark that we got rid of one of the diverging terms this way as 
$\alpha_{\rm reg} = 2 \alpha^{(-1,-1)}+ \alpha^{(-2)} = \frac{9+4h}{27} + \mathcal{O}(\epsilon) $. 
At this point, we have no choice but to admit that there exists another field in the 
theory with dimension $3$ at $c=-2$ to cancel the last diverging term. As we already 
discussed, this field has to be $\Phi_{1,7}$ with conformal weight 
$h_{\psi} \equiv h_{1,7} = 3 + 3 \epsilon + \mathcal{O}(\epsilon^2)$. The OPE thus reads
\begin{align}
\displaystyle \Phi_h(z) \Phi_h(0) \sim \frac{a_\Phi}{z^{2h-1}} \left[  z^{h_{\xi} - 1} \right. & \xi(0)  +\frac{z^{h_{\xi}}}{2} \partial \xi(0) + \alpha^{(-1,-1)} z^{h_{\xi} + 1} \phi(0) \notag \\
&   \left. +  \alpha_{\rm reg} z^{h_{\xi} + 1} L_{-2} \xi(0) + z^{h_{\psi} - 1} \Phi_{1,7} (0) +\dots \right] .
\end{align}
We define the field $\psi(z)$ as
\begin{equation}
\displaystyle  \Phi_{1,7} (z)  = \frac{\alpha^{(-1,-1)} \Braket{\phi | \phi}}{\beta(\epsilon)} \psi (z) - \alpha^{(-1,-1)} \phi(z),
\end{equation}
where $\beta = - \frac{ \Braket{\phi | \phi}}{h_{\psi}-h_{\xi}-2}$. We finally 
notice that $\alpha^{(-1,-1)} (h_{\psi}-h_{\xi}-2)=-\frac{h}{9}+\mathcal{O}(\epsilon)$, 
so that the OPE becomes regular at $c=-2$ and involves logarithms
\begin{equation}
\displaystyle \Phi_h(z) \Phi_h(0) \sim \frac{a_\Phi}{z^{2h-1}} \left[ \xi(0) +\frac{z}{2} \partial \xi(0) + \frac{9+4h}{27} z^{2} L_{-2} \xi(0) + \frac{h}{9} z^2 (\psi (0) +  \phi (0) \log z) +\dots \right].
\end{equation}
One can check that the operators $\psi$ and $\phi$ defined this way satisfy the 
usual OPEs for logarithmic operators. In particular, it is straightforward to show that 
\begin{equation}
\displaystyle \left\langle \phi(z) \psi(0)\right\rangle = \frac{\beta_{1,7}}{z^6},
\end{equation}
where 
\begin{equation}
\displaystyle \beta_{1,7} = \lim_{\epsilon \rightarrow 0} \beta (\epsilon) = - \lim_{\epsilon \rightarrow 0} \frac{ \Braket{\phi | \phi}}{h_{\psi}-h_{\xi}-2} = - 18  ,
\end{equation}
as $\Braket{\phi | \phi} = 27 \epsilon + \mathcal{O}(\epsilon^2)$. We find the 
same $\beta$ parameter as \cite{KauschGaberdiel} but with a different (technically simpler, but less rigorous) argument 
which only involves computation of a few Virasoro commutators.

We now turn to a more general LCFT with central charge $c=1-\frac{6}{x_0 (x_0+1)}$. 
Using the previous results, we conjecture that the indecomposability parameter $\beta$ 
for a generic Jordan cell with structure~\eqref{eq_general_proj}, can be computed from 
small deformations around this theory $x=x_0+ \epsilon$ as
\begin{equation}
\label{b_formula}
\displaystyle \boxed{\beta = - \lim_{\epsilon \rightarrow 0} \frac{ \Braket{\phi | \phi}}{h_{\psi}-h_{\xi}-n} = -  \frac{\left. \frac{\rm d}{\mathrm{d} \epsilon} \Braket{\phi | \phi}\right|_{\epsilon=0}  }{\left. \frac{\rm d}{\mathrm{d} \epsilon} (h_{\psi}-h_{\xi}) \right|_{\epsilon=0}}   } ,
\end{equation}
where $n=(\left. h_{\psi}-h_{\xi})\right|_{\epsilon=0}$.
We will show in the 
two next sections that eq.~\eqref{b_formula} is consistent with the previous 
studies and agrees very well with numerical results. The problem now reduces to identifying Jordan 
cells of a given theory. This problem is fairly well understood, and we shall now turn 
to concrete lattice examples to illustrate this.


\section{Lattice models and algebraic considerations}

Although the systematic study of Virasoro indecomposable modules can be 
analyzed on the Virasoro side in an rather abstract way \cite{KauschGaberdiel,KytolaRidout}, 
it is also instructive to analyze how indecomposability arises directly from lattice 
regularizations \cite{RS3,PRZ}.
The structure of the Virasoro staggered modules can be predicted from the analysis 
of the projective modules of associative lattice algebras such as the Temperley-Lieb 
algebra \cite{RS3, RP1,RP2}. These results are of course consistent with those of the 
Virasoro-based approach. This section follows closely the results of Ref. \cite{RS3}.

\subsection{`Dense' LCFTs from Temperley-Lieb algebra, XXZ spin-1/2 chain, dense loops and supersymmetry}

\begin{figure}
\begin{center}
\begin{pspicture}(0,0)(9,6)
\rput[Bc](1.0,0.5){$\bar{\square} \otimes \square$}
 \psline[linecolor=black,linewidth=0.5pt,arrowsize=5pt]{->}(0,1.0)(1.0,2.0) 
 \psline[linecolor=black,linewidth=0.5pt,arrowsize=5pt]{->}(1.0,2.0)(0.0,3.0) 
 \psline[linecolor=black,linewidth=0.5pt,arrowsize=5pt]{->}(1.0,2.0)(2.0,1.0) 
 \psline[linecolor=black,linewidth=0.5pt,arrowsize=5pt]{->}(2.0,3.0)(1.0,2.0) 
 \rput[Bc](3.0,2.0){$=$}
 \psellipticarc[linecolor=red,linewidth=1.0pt]{->}(6.0,2.0)(0.71,0.71){135}{225}
 \psellipticarc[linecolor=red,linewidth=1.0pt]{->}(4.0,2.0)(0.71,0.71){315}{45}
  \rput[Bc](5.0,0.5){$p_B$}
   \rput[Bc](6.5,2.0){$+$}
 \psellipticarc[linecolor=red,linewidth=1.0pt]{<-}(8.0,1.0)(0.71,0.71){45}{135}
 \psellipticarc[linecolor=red,linewidth=1.0pt]{<-}(8.0,3.0)(0.71,0.71){225}{315}   
 \rput[Bc](8.0,0.5){$1-p_B$}
 
 \rput[Bc](1.0,3.5){$\square \otimes \bar{\square}$}
 \psline[linecolor=black,linewidth=0.5pt,arrowsize=5pt]{<-}(0,4.0)(1.0,5.0) 
 \psline[linecolor=black,linewidth=0.5pt,arrowsize=5pt]{<-}(1.0,5.0)(0.0,6.0) 
 \psline[linecolor=black,linewidth=0.5pt,arrowsize=5pt]{<-}(1.0,5.0)(2.0,4.0) 
 \psline[linecolor=black,linewidth=0.5pt,arrowsize=5pt]{<-}(2.0,6.0)(1.0,5.0) 
 \rput[Bc](3.0,5.0){$=$}
 \psellipticarc[linecolor=red,linewidth=1.0pt]{<-}(6.0,5.0)(0.71,0.71){135}{225}
 \psellipticarc[linecolor=red,linewidth=1.0pt]{<-}(4.0,5.0)(0.71,0.71){315}{45}
  \rput[Bc](5.0,3.5){$1-p_A$}
   \rput[Bc](6.5,5.0){$+$}
 \psellipticarc[linecolor=red,linewidth=1.0pt]{->}(8.0,4.0)(0.71,0.71){45}{135}
 \psellipticarc[linecolor=red,linewidth=1.0pt]{->}(8.0,6.0)(0.71,0.71){225}{315}   
 \rput[Bc](8.0,3.5){$p_A$}
\end{pspicture}
\end{center}
  \caption{Graphical representation of the Temperley-Lieb-based vertex model. The 
  lattice consists of alternating arrows going up for $i$ even and down for $i$ odd, 
  where $i=1,\dots,L=2N$ corresponds to the horizontal (space) coordinate. The system 
  has free boundary conditions in the horizontal direction and periodic in the vertical 
  (imaginary time) direction. We choose each vertex according to its probability; this 
  draws a dense loop configuration on the lattice. Each closed loop carries a weight $n=q+q^{-1}$. 
   In the supersymmetric language, the alternating $\square,\bar{\square}$ representations 
   correspond to a lattice orientation, conserved along each loop. The system is isotropic 
   when $p_A = p_B$, while the transition occurs when $p_A = 1-p_B$.}
  \label{figTL}
\end{figure}
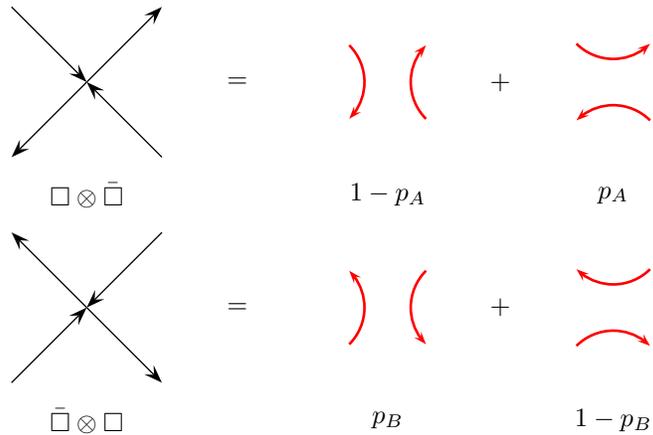

\subsubsection{Temperley-Lieb algebra}

Underlying most of the models we shall consider is the Temperley-Lieb algebra 
$TL_{2N}(q)$. The algebra $TL_{2N}(q)$ defined on $L=2N$ ($N \in \mathbb{N}/2$) 
strands consists of all the words written with the $2N-1$ generators $e_i$ ($1 \leq i \leq 2N-1$), subject to the relations
\begin{subequations} \label{TLdef}
\begin{eqnarray}
\left[ e_i , e_j \right] &=&0 \ (\left|i-j \right| \geq 2 )\\
e_i ^2 &=& n e_i\\
e_i e_{i \pm 1} e_i &=& e_i
\end{eqnarray}
\end{subequations}
with
\begin{equation}
\displaystyle{ n=q+q^{-1}}.
\end{equation}
We also define $q=\mathrm{e}^{i \gamma}$ and $\gamma = \frac{\pi}{x+1}$.
The action of $e_i$ as an operator acting on $2N$ strands can be naturally 
represented graphically. This constructs what we shall refer to as loop or adjoint representation.
The standard modules of the TL algebra are well known, and have dimensions
\begin{equation}
\displaystyle{d_j = \left( \begin{array}{c} 2N  \\ N + j  \end{array} \right) -  \left( \begin{array}{c} 2N  \\ N + j +1 \end{array} \right)},
\end{equation}
where in the geometrical language, $2j$ is the number of `through lines' or `strings' 
that propagate along the imaginary time direction. We have $j \in \mathbb{N}$, restricted 
to the condition $N + j \in \mathbb{N}$, so if $N$ is a half-integer ($L=2N$ odd), so must 
be $j$. These standard modules are irreducible for $q$ generic ({\it i.e.} not a root of unity).

We consider a two-dimensional model defined by the transfer matrix
\begin{equation}
\displaystyle{T = \prod_{i=1}^{N-1} \left( p_B + (1-p_B) \ e_{2i} \right) \prod_{i=1}^{N} \left( (1-p_A) + p_A \ e_{2i-1} \right) } ,
\end{equation}
which acts on a given TL module. This definition is valid for $L=2N$ even 
but it can be readily adapted to an odd number of sites. We will mainly work 
with three different representations: geometrical (adjoint), 6-vertex, and supersymmetric. 
Using the geometrical (loop) representation of $TL_{2N}(q)$, we obtain a dense loop model, 
where each closed loop carries a weight $n$ (fugacity). A graphical representation of this 
vertex model is given Fig.~\ref{figTL}.
It sometimes helps to think in terms of a Q-state Potts model whose high-temperature expansion 
consists of drawing dense loop configurations with fugacity $n=\sqrt{Q}$.

In the strong anisotropy limit $p_A \rightarrow 0$ with $p_A / (1-p_B)$ fixed, we can extract 
the Hamiltonian of the equivalent one-dimensional quantum system. It reads, up to an irrelevant constant,
\begin{equation}
\displaystyle{ H = - \varepsilon \sum_{i=1}^{N-1} e_{2i} - \varepsilon^{-1} \sum_{i=1}^{N} e_{2i-1}},
\end{equation}
where $\varepsilon = \sqrt{p_A /(1-p_B)}$. The system is isotropic when $p_A = p_B$, while the transition 
occurs when $p_A = 1-p_B$. Hereafter, we will always consider the case $\varepsilon = 1$. 

Finally, we recall some useful relations for this integrable model. Using Bethe ansatz, one can show 
\cite{PottsBethe} that the mean value of the TL generators on the groundstate is 
\begin{equation}
\displaystyle e_{\infty} = \sin^2 \gamma \int_{-\infty}^{+\infty} \frac{{\rm d} x}{\cosh \pi x} \ \frac{1}{\cosh 2 \gamma x - 2 \cos \gamma}.
\label{e_inf}
\end{equation}
Meanwhile, the dispersion relation of the quasi-particles is $\varepsilon \sim v_F |k - k_F|$, where the sound velocity reads
\begin{equation}
\displaystyle v_{F} = \frac{\pi \sin \gamma}{\gamma}.
\end{equation}

\subsubsection{6-vertex or XXZ spin chain representation}

Another natural representation of the Temperley-Lieb algebra is provided by the 
6-vertex model. We write $n = \mathrm{e}^{i \gamma} + \mathrm{e}^{-i \gamma}$ and 
$q = \mathrm{e}^{i \gamma}$. The Hamiltonian limit of the 6-vertex model is the XXZ 
chain, with Hilbert space $ \mathcal{H}_{\rm XXZ} = (\mathbb{C}^2)^{\otimes 2N} $. We will 
focus on this limit hereafter. The Temperley-Lieb generators in this representation read
\begin{equation}
e_i = \mathbb{I} \otimes \mathbb{I} \otimes \dots \otimes \left( \begin{array}{cccc}
0 & 0 & 0 & 0 \\
0 & q^{-1} & -1 & 0 \\
0 & -1 & q & 0 \\
0 & 0 & 0 & 0 \end{array} \right) \otimes \dots \otimes \mathbb{I},
\label{e_eiXXZ}
\end{equation}
where we have used the basis $\lbrace \Ket{\uparrow \uparrow}, \Ket{\uparrow \downarrow}, \Ket{\downarrow \uparrow}, \Ket{\downarrow \downarrow} \rbrace$ of $ (\mathbb{C}^2)^{\otimes 2}$, the tensor product of the Hilbert spaces 
of the sites $i$ and $i+1$. We can check that the generators indeed satisfy the TL algebra, 
with $n =  q + q^{-1}$. We can express the Hamiltonian of the XXZ chain in terms of the usual Pauli operators
\begin{equation}
\displaystyle H = \frac{1}{2} \sum_{i=1}^{2N-1} \left( \sigma^x_i \sigma^x_{i+1} + \sigma^y_i \sigma^y_{i+1}  + \frac{q + q^{-1}}{2} \sigma^z_i \sigma^z_{i+1}  \right) + \frac{q - q^{-1}}{4} \left( \sigma^z_1 - \sigma^z_{2N} \right) - N \cos \gamma.
\end{equation}
Note that this XXZ Hamiltonian (or equivalently the transfer matrix of the 6-Vertex model) 
commutes \cite{PasquierSaleur} with the generators of the quantum group $U_q(\mathfrak{sl}_2)$.

\subsubsection{Supersymmetric representation (SUSY)}
We introduce in this paragraph a supersymmetric formulation of our model, which provides 
another natural representation of $TL_{2N}(q)$ \cite{RS1}. We consider that each edge of 
our two-dimensional lattice carries a $\mathbb{Z}_2$ graded vector space of dimension $n+m|m$. 
We choose these vector spaces to be the fundamental $\square$ of the Lie superalgebra 
$\mathfrak{gl}(n+m|m)$ for $i$ odd (corresponding to down arrows of Fig.~\ref{figTL}), 
and the dual $\bar{\square}$ for $i$ even (up arrows). The transfer matrix (or the Hamiltonian) 
then acts on the graded tensor product $\mathcal{H} = (\square \otimes \bar{\square})^{\otimes N}$. 
The TL generators are simply quadratic Casimir invariants,  providing a natural generalization 
of the Heisenberg chain to the $\mathfrak{gl}(n+m|m)$ algebra. We can check that a loop expansion 
of the transfer matrix yields a dense loop model with a weight $\mathrm{str} \ \mathbb{I} = n + m -m = n$ 
for each closed loop as expected.
There is a continuum quantum field theory associated with this spin chain, which turns out 
\cite{RS1} to be a non-linear $\sigma$-model on complex projective space 
$\mathbb{CP}^{n+m-1|m} = \mathrm{U}(m+n|m) / (\mathrm{U}(1) \times \mathrm{U}(m+n-1|m)) $ at 
topological angle $\theta = \pi$.

\subsubsection{Continuum limit: the generic case}
Let us also say a few words about the continuum limit. The XXZ chain with 
$q=\mathrm{e}^{i \gamma}$, $\gamma = \pi / (x+1)$ is described by a CFT 
\cite{PasquierSaleur} with central charge
\begin{equation}
\displaystyle c = 1 - \frac{6}{x (x+1)}.
\end{equation}
Consider the Verma modules $\mathcal{V}_{r,s}$, spanned by the action of 
the Virasoro generators $L_n$ with $n<0$ on the highest weight state with 
conformal dimension $h_{r,s}$ given by the Kac formula 
\begin{equation}
\displaystyle h_{r,s} = \frac{ \left[(x+1)r - xs \right]^2 - 1}{4 x (x+1)}.
\end{equation}
When we take the scaling limit of the XXZ chain, the conformal weights that 
occur in the spectrum are $h_{1,1+2j}$ in the Kac table, where $j=n, \ n \in \mathbb{N}$ 
if L even and $j=n/2, \ n \in 2\mathbb{N}+1$ if L odd. For $q$ generic (not a root of unity), 
there is a single null vector in the Verma module $\mathcal{V}_{1,1+2j}$ at conformal weight 
$h_{1,-1-2j}$ that must be set to zero in order to obtain a simple (irreducible) module. 
Hence, we define the standard (Kac) modules $r_j = r_{1,1+2j} = \mathcal{V}_{1,1+2j} / \mathcal{V}_{1,-1-2j} $ 
which are irreducible for $q$ generic. The character\footnote{The modular parameter $q$ in 
characters and partition functions has of course nothing to do with $q=\mathrm{e}^{i \gamma}$ 
that parametrizes the weight of closed loops in the Temperley-Lieb algebra. We keep the same 
notations and hope this will not confuse the reader.} of the module $r_j$ reads
\begin{equation}
\displaystyle K_j = \mathrm{Tr}_{r_j} \ q^{L_0-c/24} =  \frac{q^{h_{1,1+2j}-c/24}-q^{h_{1,-1-2j}-c/24} }{P(q)}.
\end{equation}
where $P(q)$ is the inverse of the Euler partition function, and is related to the Dedekind $\eta$ function
\begin{equation}
\displaystyle P(q) =  \prod_{n=1}^{\infty} (1 - q^n) = q^{-1/24} \eta (q).
\end{equation}
The partition function of the sector of spin $S_z$ reads
\begin{equation}
\displaystyle Z^{\mathrm{XXZ}}_{S_z} = \mathrm{Tr} \ q^{L_0-c/24} = \sum_{j=|S_z|}^{\infty} K_j ,
\end{equation}
the global partition function of the $U_q(\mathfrak{sl}_2)$ XXZ spin chain is then readily obtained
\begin{equation}
\displaystyle Z^{\mathrm{XXZ}} = \sum_{S_z=-\infty}^{+\infty} Z^{\mathrm{XXZ}}_{S_z} = \sum_{j=0}^{\infty} (2 j + 1) K_j .
\end{equation}
The partition function of the superspin chain with $\mathfrak{gl}(n+m|n)$ symmetry is given 
by a similar expression where one replaces $2j+1$ by the dimensions of the irreducible 
representations of the commutant $\mathcal{A}_{n+m|n}(2N)$ of $TL_{2N}(q)$ in the SUSY 
representation \cite{RS2}. The algebra $\mathcal{A}_{n+m|n}$  is in fact much larger than $\mathfrak{gl}(n+m|n)$.

These results were obtained for $q$ generic but are supposed to remain correct even 
when $q$ is a root of unity even though the Virasoro standard modules are no longer 
irreducible in general.

\subsection{`Dilute' LCFTs from the integrable $O(n)$ model}

We will also consider a fundamentally different version of the previous LCFTs using 
the integrable dilute $O(n)$ on the square lattice. We shall refer to these theories 
as `dilute' as opposed to the `dense' ones based on the Temperley-Lieb algebra. This 
denomination obviously refers to the dense or dilute nature of the underlying loop gas.
Therefore, we describe in this section the $O(n)$ model defined on an annulus of 
width $2N$. It corresponds to a dilute loop model where closed loops carry a weight 
$n$; we shall focus here only on the dilute phase. This model also possesses a dense phase 
which is in the same universality class as the dense loop model. Note that the case $n \rightarrow 0$ 
is obviously relevant for the physics of polymers. In terms of spin chains, it is described 
by a $S=1$ $U_q(\mathfrak{sl}_2)$-invariant chain where the states $S_z=\pm 1$ are viewed as occupied 
by parts of loops and $S_z= 0$ as empty. This model also corresponds to $\mathfrak{osp}(n+2m|2m)$ 
(super)spin chains and to non-linear sigma models with supersphere target space $S^{2m+n-1|2m} \simeq \mathrm{OSp}(2m+n|2m)/\mathrm{OSp}(2m+n-1|2m)$ \cite{RS1}. There is a dilute version of the Temperley-Lieb algebra behind 
all these models. We will not go into the details of these different formulations here 
but only describe the geometrical setup that will allow us to measure indecomposability 
parameters in the next section.

\subsubsection{Lattice model}

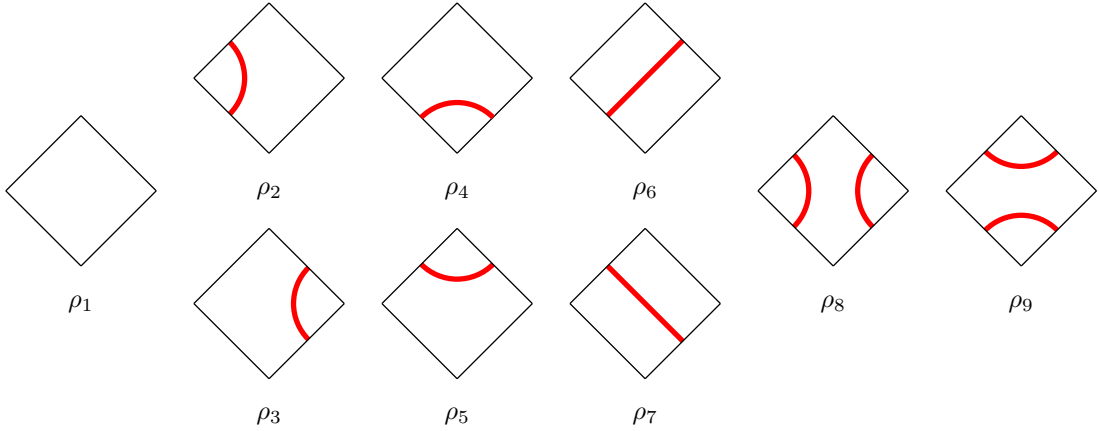
\begin{figure}
\begin{center}
\begin{pspicture}(-0.5,-0.5)(14,5.5)
 \psline[linecolor=black,linewidth=0.5pt,arrowsize=5pt]{-}(-0.5,3)(0.5,4) 
 \psline[linecolor=black,linewidth=0.5pt,arrowsize=5pt]{-}(0.5,4)(1.5,3) 
 \psline[linecolor=black,linewidth=0.5pt,arrowsize=5pt]{-}(-0.5,3)(0.5,2) 
 \psline[linecolor=black,linewidth=0.5pt,arrowsize=5pt]{-}(0.5,2)(1.5,3)
  \rput[Bc](0.5,1.5){$\rho_1$}
 
 \psellipticarc[linecolor=red,linewidth=2.0pt]{-}(2,4.5)(0.71,0.71){315}{45}  
 \psline[linecolor=black,linewidth=0.5pt,arrowsize=5pt]{-}(2,4.5)(3,5.5) 
 \psline[linecolor=black,linewidth=0.5pt,arrowsize=5pt]{-}(3,5.5)(4,4.5) 
 \psline[linecolor=black,linewidth=0.5pt,arrowsize=5pt]{-}(2,4.5)(3,3.5) 
 \psline[linecolor=black,linewidth=0.5pt,arrowsize=5pt]{-}(3,3.5)(4,4.5)
 \rput[Bc](3.0,3.0){$\rho_2$}

 \psellipticarc[linecolor=red,linewidth=2.0pt]{-}(4,1.5)(0.71,0.71){135}{225} 
 \psline[linecolor=black,linewidth=0.5pt,arrowsize=5pt]{-}(2,1.5)(3,2.5) 
 \psline[linecolor=black,linewidth=0.5pt,arrowsize=5pt]{-}(3,2.5)(4,1.5) 
 \psline[linecolor=black,linewidth=0.5pt,arrowsize=5pt]{-}(2,1.5)(3,0.5) 
 \psline[linecolor=black,linewidth=0.5pt,arrowsize=5pt]{-}(3,0.5)(4,1.5)
  \rput[Bc](3.0,0.0){$\rho_3$}

 \psellipticarc[linecolor=red,linewidth=2pt]{-}(5.5,2.5)(0.71,0.71){225}{315} 
  \psline[linecolor=black,linewidth=0.5pt,arrowsize=5pt]{-}(4.5,4.5)(5.5,5.5) 
 \psline[linecolor=black,linewidth=0.5pt,arrowsize=5pt]{-}(5.5,5.5)(6.5,4.5) 
 \psline[linecolor=black,linewidth=0.5pt,arrowsize=5pt]{-}(4.5,4.5)(5.5,3.5) 
 \psline[linecolor=black,linewidth=0.5pt,arrowsize=5pt]{-}(5.5,3.5)(6.5,4.5)
\rput[Bc](5.5,3.0){$\rho_4$}

   \psellipticarc[linecolor=red,linewidth=2pt]{-}(5.5,3.5)(0.71,0.71){45}{135}
 \psline[linecolor=black,linewidth=0.5pt,arrowsize=5pt]{-}(4.5,1.5)(5.5,2.5) 
 \psline[linecolor=black,linewidth=0.5pt,arrowsize=5pt]{-}(5.5,2.5)(6.5,1.5) 
 \psline[linecolor=black,linewidth=0.5pt,arrowsize=5pt]{-}(4.5,1.5)(5.5,0.5) 
 \psline[linecolor=black,linewidth=0.5pt,arrowsize=5pt]{-}(5.5,0.5)(6.5,1.5)
   \rput[Bc](5.5,0.0){$\rho_5$}

 \psline[linecolor=red,linewidth=2pt]{-}(7.5,4)(8.5,5)   
   \psline[linecolor=black,linewidth=0.5pt,arrowsize=5pt]{-}(7,4.5)(8,5.5) 
 \psline[linecolor=black,linewidth=0.5pt,arrowsize=5pt]{-}(8,5.5)(9,4.5) 
 \psline[linecolor=black,linewidth=0.5pt,arrowsize=5pt]{-}(7,4.5)(8,3.5) 
 \psline[linecolor=black,linewidth=0.5pt,arrowsize=5pt]{-}(8,3.5)(9,4.5)
\rput[Bc](8,3.0){$\rho_6$}

 \psline[linecolor=red,linewidth=2pt]{-}(7.5,2)(8.5,1)  
 \psline[linecolor=black,linewidth=0.5pt,arrowsize=5pt]{-}(7,1.5)(8,2.5) 
 \psline[linecolor=black,linewidth=0.5pt,arrowsize=5pt]{-}(8,2.5)(9,1.5) 
 \psline[linecolor=black,linewidth=0.5pt,arrowsize=5pt]{-}(7,1.5)(8,0.5) 
 \psline[linecolor=black,linewidth=0.5pt,arrowsize=5pt]{-}(8,0.5)(9,1.5)
   \rput[Bc](8,0.0){$\rho_7$}

 \psellipticarc[linecolor=red,linewidth=2.0pt]{-}(11.5,3.0)(0.71,0.71){135}{225}
 \psellipticarc[linecolor=red,linewidth=2.0pt]{-}(9.5,3.0)(0.71,0.71){315}{45}
 \psline[linecolor=black,linewidth=0.5pt,arrowsize=5pt]{-}(9.5,3)(10.5,4) 
 \psline[linecolor=black,linewidth=0.5pt,arrowsize=5pt]{-}(10.5,4)(11.5,3) 
 \psline[linecolor=black,linewidth=0.5pt,arrowsize=5pt]{-}(9.5,3)(10.5,2) 
 \psline[linecolor=black,linewidth=0.5pt,arrowsize=5pt]{-}(10.5,2)(11.5,3)
  \rput[Bc](10.5,1.5){$\rho_8$}

 \psellipticarc[linecolor=red,linewidth=2pt]{-}(13.0,2.0)(0.71,0.71){45}{135}
 \psellipticarc[linecolor=red,linewidth=2pt]{-}(13.0,4.0)(0.71,0.71){225}{315} 
 \psline[linecolor=black,linewidth=0.5pt,arrowsize=5pt]{-}(12,3)(13,4) 
 \psline[linecolor=black,linewidth=0.5pt,arrowsize=5pt]{-}(13,4)(14,3) 
 \psline[linecolor=black,linewidth=0.5pt,arrowsize=5pt]{-}(12,3)(13,2) 
 \psline[linecolor=black,linewidth=0.5pt,arrowsize=5pt]{-}(13,2)(14,3)
  \rput[Bc](13,1.5){$\rho_9$}
          
\end{pspicture}
\end{center}
  \caption{Plaquettes and integrable weights for the dilute $O(n)$ model on the square lattice}
  \label{figOn}
\end{figure}
Our starting point is the expression of the integrable version of the 
dilute $O(n)$ model defined on the square lattice. The $\v{R}$ matrix is the 
sum of 9 plaquettes represented graphically in Figure \ref{figOn}
\begin{equation}
\displaystyle \v{R}_j(u) = \sum_{i=1}^{9} \rho_i(u) \mathcal{O}_j^i,
\end{equation}
where $j \in \{1, \dots, 2N-1 \}$ labels the sites.
It satisfies the Yang-Baxter equations for the corresponding integrable weights \cite{Onweights}
\begin{subequations}
\begin{eqnarray}
\rho_1 &=& 1+\frac{\sin u \sin (3 \lambda - u)}{\sin 2 \lambda \sin 3 \lambda} \\
\rho_2 &=& \rho_3 = \frac{\sin (3 \lambda - u)}{\sin 3 \lambda}\\
\rho_4 &=& \rho_5 = \frac{\sin u}{\sin 3 \lambda}\\
\rho_6 &=& \rho_7 = \frac{\sin u \sin(3 \lambda -u)}{\sin 2 \lambda \sin 3 \lambda}\\
\rho_8 &=& \frac{\sin (2 \lambda -u) \sin(3 \lambda -u)}{\sin 2 \lambda \sin 3 \lambda}\\
\rho_9 &=& -\frac{\sin u \sin(\lambda -u)}{\sin 2 \lambda \sin 3 \lambda}
\end{eqnarray}
\end{subequations}
where $n = - 2 \cos 4 \lambda$ is the weight given to every closed loop. Note that
\begin{equation}
\displaystyle \v{R}_i(u=0) =  \mathcal{O}_i^1 + \mathcal{O}_i^2 + \mathcal{O}_i^3 + \mathcal{O}_i^8 = \mathrm{Id}.
\end{equation}
One can extract the corresponding 1D Hamiltonian $H= - \left. \frac{\mathrm{d}\v{R}}{\mathrm{d}u}\right|_{u=0}$ using an expansion in the 
spectral parameter $u$. We find that the interaction between the sites $i$ and $i+1$ reads
\begin{align}
\displaystyle h_{i,i+1} = \frac{1}{\sin 2 \lambda} \mathcal{O}_i^1 & - \mathrm{cotan} 3 \lambda (\mathcal{O}_i^2 + \mathcal{O}_i^3) + \frac{1}{ \sin 3 \lambda} (\mathcal{O}_i^4 + \mathcal{O}_i^5) \notag \\
  +& \frac{1}{ \sin 2 \lambda} (\mathcal{O}_i^6 + \mathcal{O}_i^7) - (\mathrm{cotan} 2 \lambda + \mathrm{cotan} 3 \lambda) \mathcal{O}_i^8 - \frac{\sin \lambda}{\sin 2 \lambda \sin 3 \lambda} \mathcal{O}_i^9.
 \label{hamOn}
\end{align}
In order to obtain an integrable Hamiltonian in the case of open (free) boundary 
conditions, we would need to consider K matrices at the edges \cite{Skyanin}. Instead, 
we here consider the simpler (non-integrable) Hamiltonian $H = -\sum h_{i,i+1}$. In what follows, every 
reference to the $O(n)$ model corresponds to this Hamiltonian. 
The Fermi velocity and the mean value of $h_{i,i+1}$ on the ground state are \cite{OnBethe}
\begin{equation}
\displaystyle v_F = \frac{\pi}{3 \lambda},
\end{equation}
and
\begin{equation}
\displaystyle h_{\infty} = 2 \int_{-\infty}^{+\infty} {\rm d} k \ \frac{\tanh 3 \lambda k \cosh (5 k \lambda - k \pi) \cosh k \lambda}{\sinh k \pi}.
\end{equation}

\subsubsection{Continuum limit}

The CFT describing the scaling limit of dilute loops has central charge \cite{Onc}
\begin{equation}
\displaystyle c = 1-3 \frac{(4 \lambda - \pi)^2}{2 \lambda \pi},
\end{equation}
with $n= - 2 \cos 4 \lambda$ and $\lambda \in [\pi/4,\pi/2]$. The conformal weights 
appearing in the spectrum lie now in the first row of the Kac table $h_{1+2j,1}$.
The trace over the standard module with $2j$ through lines reads
\begin{equation}
\displaystyle K_{j} = \mathrm{Tr}_{r_j} \ q^{L_0-c/24} = \frac{q^{h_{1+2j,1}-c/24}-q^{h_{-1-2j,1}-c/24} }{P (q)}.
\end{equation}
As in the dense case, the partition function of the spin-1 XXZ chain is just the sum 
of these characters with a degeneracy $2j+1$. In the case of $\mathfrak{osp}(n+2m|2m)$-invariant 
superspin chains, the degeneracy $2j+1$ is replaced by the dimension of the corresponding irreducible 
representation of the commutant $\mathcal{B}_{n+2m|2m}(2N)$ of the dilute Temperley-Lieb algebra \cite{RS2}.

\subsection{Indecomposability and lattice Jordan cells at $q$ root of unity on a concrete example: the logarithmic Ising model $\mathcal{LM}(3,4)$}

\label{subsec_ising}

We now turn to the study of indecomposability when $q$ is a root of unity. In this case, 
the standard modules of the Temperley-Lieb algebra still exist but may no longer be irreducible. 
Read and Saleur \cite{RS3} showed in details that the structure of the projective modules of 
the Temperley-Lieb algebra parallels that of several staggered Virasoro modules in the continuum 
limit. The general algebraic structure of the XXZ spin chain was illustrated in terms of 
`staircase diagrams' as a representation (bimodule) of $TL_{2N}(q) \otimes U_q(\mathfrak{sl}_2)$, or 
$TL_{2N}(q) \otimes \mathcal{A}_{n+m|n}(2N)$ in the SUSY case.
In this paper, we will pass directly to the continuum theory and only mention the important 
algebraic lattice results when necessary. We refer the interested reader to Ref.~\cite{RS3} for more details.

Let us consider the case $q=\mathrm{e}^{i \pi/4}$, which corresponds to a dense CFT with 
central charge $c=\frac{1}{2}$. Loop expansion of the partition function yields a loop 
model with fugacity $n=\sqrt{2}$; this is of course the celebrated (logarithmic) Ising model $\mathcal{LM}(3,4)$ \cite{PRZ}.
The Kac formula at $x=3$ reads
\begin{equation}
\displaystyle h_{r,s} = \frac{ \left[4r - 3 s \right]^2 - 1}{48},
\end{equation}
and the values appearing in the spectrum of the $U_q(\mathfrak{sl}_2)$-invariant XXZ 
spin chain at $q=\mathrm{e}^{i \pi/4}$ are
\begin{equation}
\displaystyle h_{1,1+2j} = \frac{j(3j-1)}{4}.
\end{equation}
We will focus the continuum limit of a spin chain with an even number of sites so 
that $j \in \mathbb{N}$; the case $j = n/2, \ n \in 2\mathbb{N}+1$ is treated in 
a similar way. Note that we have $h_{r,s}=h_{-r,-s}=h_{r+3,s+4}$. The character 
of the standard module $r_j$ reads
\begin{equation}
\displaystyle K_{j} = \mathrm{Tr}_{r_j} \ q^{L_0-c/24} = \frac{q^{(1- 6 j)^2/48}-q^{(7+6j)^2/48} }{\eta (q)},
\end{equation}
whereas the characters \cite{Characters} of the simple (irreducible) Virasoro module are
\begin{subequations}
\begin{eqnarray}
\chi_{j=4p}&=&
\sum_{n\notin [-2p,-1]}
\frac{q^{(24n-1+24p)^{2}/48}-q^{(24n+17+24p)^{2}/48}}
{\eta(q)},
\\
\chi_{j=4p+1}&=&
\sum_{n\notin [-2p,-1]}
\frac{q^{(24n+5+24p)^{2}/48}-q^{(24n+11+24p)^{2}/48}}
{\eta(q)},
\\
\chi_{j=4p+2}&=&
\sum_{n\notin [-2p-1,-1]}
\frac{q^{(24n+11+24p)^{2}/48}-q^{(24n+29+24p)^{2}/48}}
{\eta(q)},
\\
\chi_{j=4p+3}&=&
\sum_{n\notin [-2p-1,-1]}
\frac{q^{(24n+17+24p)^{2}/48}-q^{(24n+23+24p)^{2}/48}}
{\eta(q)}.
\end{eqnarray}
\end{subequations}
We can see using these character identities that the standard 
modules $r_j$, $j \in \mathbb{N}$ are no longer irreducible.
The decomposition onto the simple characters indeed yields
\begin{subequations}
\begin{eqnarray}
K_{j=4p}&=&\chi_{4p}+\chi_{4p+3}, \\
K_{j=4p+1}&=&\chi_{4p+1}+\chi_{4p+2}, \\
K_{j=4p+2}&=&\chi_{4p+2}+\chi_{4p+5}, \\
K_{j=4p+3}&=&\chi_{4p+3}+\chi_{4p+4}.
\end{eqnarray}
\end{subequations}
Let $R_{j}$ be the simple Virasoro module with conformal weight 
$h_{1,1+2j}$ and character $\chi_j$.
From lattice algebraic considerations, we know that the standard 
modules $r_j$ must thus have the structure
\begin{equation}
r_{j}=~~~~~\left\{\begin{array}{cl}
\begin{array}{ccc}
R_{j}&&\\
&\hskip-.2cm\searrow&\\
&&\hskip-.3cmR_{j+3}
\end{array} &\hbox{$j\equiv0 \ ($mod$ 4)$ or $j\equiv2 \ ($mod$ 4)$}\nonumber\\

\begin{array}{ccc}
R_{j}&&\\
&\hskip-.2cm\searrow&\\
&&\hskip-.3cmR_{j+1}
\end{array} &\hbox{$j\equiv1 \ ($mod$ 4)$ or $j\equiv3 \ ($mod$ 4)$}\nonumber\\
\end{array}\right.
\end{equation}
The arrows represent again the action of the Virasoro algebra. 
The bottom submodule is invariant while it is possible to go from 
the top to the bottom acting by some element of Virasoro. The top 
simple quotient must have a smaller $j$ number than the bottom \cite{RS3}.

Using the standard modules as elementary bricks, it becomes quite 
easy to construct the staggered modules of the theory using the knowledge, again from the lattice, 
that they must be diamond-shaped. For instance for $j=2$, we expect
\begin{equation}
\mathcal{P}_2=
 \begin{array}{ccccc}
      &&\hskip-.7cm R_2&&\\
      &\hskip-.2cm\swarrow&\searrow&\\
      R_1&&&\hskip-.3cm R_5 \\
      &\hskip-.2cm\searrow&\swarrow&\\
      &&\hskip-.7cm R_2&&
\end{array}
\label{eq_projisingex}
\end{equation}
with character $P_2=2\chi_{2}+\chi_{1}+\chi_{5}$. This is the only 
gluing of standard modules that respects the fact that the conformal 
weights must be increasing to the right. For a given theory, 
there is always a unique way 
to construct a given diamond-shaped module in terms of simple modules\footnote{To be more precise, we mean here that 
for a given module $P_j$ of a given theory (dense or dilute), 
the values of the subscripts j of the four simples $R_j$ in the diamond $P_j$ are uniquely fixed.}
using only the structure the standard modules.

Using this method, it becomes quite straightforward to guess the general 
structure of the Virasoro staggered modules appearing in the theory 
\begin{equation}\label{eq_proj_ising}
{\cal P}_{j}=~~~~~\left\{\begin{array}{cl}
\begin{array}{ccc}
R_0&&\\
&\hskip-.2cm\searrow&\\
&&\hskip-.3cm R_3\end{array}&\hbox{$j=0$,}\\
\begin{array}{ccc}
R_1&&\\
&\hskip-.2cm\searrow&\\
&&\hskip-.3cm R_2\end{array}&\hbox{$j=1$,}\\
\begin{array}{ccccc}
     &&\hskip-.7cm R_j&&\\
     &\hskip-.2cm\swarrow&\searrow&\\
     R_{j-1}&&&\hskip-.3cm R_{j+3}\\
     &\hskip-.2cm\searrow&\swarrow&\\
     &&\hskip-.7cm R_j&&
     \end{array}&\hbox{$j\equiv0$ (mod 4) and $j>0$, or $j\equiv2$ (mod 4),}\\
     &\\
 \begin{array}{ccccc}
     &&\hskip-.7cm R_j&&\\
     &\hskip-.2cm\swarrow&\searrow&\\
     R_{j-3}&&&\hskip-.3cm R_{j+1}\\
     &\hskip-.2cm\searrow&\swarrow&\\
     &&\hskip-.7cm R_j&&
     \end{array}&\hbox{$j\equiv1$ (mod 4) and $j>1$, or $j\equiv3$ (mod 4).}\end{array}\right.
\end{equation}
Of course, it is also possible using Temperley-Lieb representation 
theory to check that a similar pattern arises from the lattice. Note 
also that it may happen that the standard module $r_j$ remains irreducible 
for $q$ a root of unity. For example, in the Ising model with $L$ odd, 
one finds that $K_{j=4p+3/2}=\chi_{4p+3/2}$ so that what we shall call staggered module 
in this case is just the simple module itself ${\cal P}_{j=4p+3/2}=r_{j=4p+3/2}=R_{j=4p+3/2}$.

Let us summarize what we have learned so far concerning the structure of 
the scaling limit of the XXZ spin chain at $q=\mathrm{e}^{i \pi/4}$ with 
an even number of sites. All the fields of the theory are organized into 
staggered modules given by~\eqref{eq_proj_ising}, $j \in \mathbb{N}$. 
Recall that each simple module $R_j$ corresponds to a field with conformal 
weight $h_{1,1+2j}$. There is a Jordan cell in $L_0$ for every staggered module
with a diamond shape. For such generic staggered module, we note the basis fields
\begin{equation}\label{eq_proj_j}
\mathcal{P}_j=
 \begin{array}{ccccc}
      &&\hskip-.7cm R_j &&\\
      &\hskip-.2cm\swarrow&\searrow&\\
      R_{j_1} &&&\hskip-.3cm R_{j_2} \\
      &\hskip-.2cm\searrow&\swarrow&\\
      &&\hskip-.7cm R_j &&
\end{array}
=
 \begin{array}{ccccc}
      &&\hskip-.7cm \psi^{(j)} &&\\
      &\hskip-.2cm\swarrow&\searrow&\\
      \xi^{(j)} &&&\hskip-.3cm \rho^{(j)} \\
      &\hskip-.2cm\searrow&\swarrow&\\
      &&\hskip-.7cm \phi^{(j)} &&
\end{array}
,
\end{equation}
with $j_1 \leq j \leq j_2$ so that  $h_{1,1+2 j_1} \leq h_{1,1+2 j} \leq h_{1,1+2 j_2}$.
We define the logarithmic coupling  $\beta_{1,1+2j}$ of this Jordan cell as
\begin{equation}
\displaystyle \beta_{1,1+2j} = \Braket{\phi^{(j)} | \psi^{(j)}}.
\end{equation}
Note that we use the same Kac labels for $\beta$ as those of the conformal weight of $\psi^{(j)}$.

We are now ready to compute this indecomposability parameter using eq.~\eqref{b_formula} 
along with $h_\xi = h_{1,1+2 j_1}$ and $h_\psi = h_{1,1+2 j_2}$.
Let us again illustrate this on a concrete example. The first Jordan cell 
occurring in the spectrum has $j=2$ and its structure is described by 
eq.~\eqref{eq_projisingex}. Defining $x=3+\epsilon$, it is straightforward 
to compute $h_{1,3}= \frac{1}{2}+\frac{\epsilon}{8} + \mathcal{O}(\epsilon^2)$ ($j=1$) 
and $h_{1,5}= \frac{5}{2}+\frac{3\epsilon}{8}+ \mathcal{O}(\epsilon^2)$ ($j=2$). 
At $\epsilon = 0$, the $L_0$ operator expressed in the basis ($\phi^{(2)}$,$\psi^{(2)}$) reads
\begin{equation}
L_0 = 
\left( \begin{array}{cc}
5/2 & 1 \\
0 & 5/2 \end{array} \right).
\end{equation}
We remark that we can readily find the relation between $\phi^{(2)}$ and 
$\xi^{(2)}$ (up to an irrelevant global normalization factor) because we 
know that $\phi^{(2)}$ must be a null-vector
\begin{equation}
\displaystyle \phi^{(2)}(z) = (L_{-2}- \frac{3}{4} L^2_{-1})\xi^{(2)}(z).
\end{equation}
In general, the operator $A$ is either known from general formulas or 
computed numerically using the Virasoro algebra. 
Straightforward commutations of the $L_{n}$'s modes yield the final result 
\begin{equation}
\label{eq_b_ising}
\displaystyle \beta_{1,5} = - \lim_{\epsilon \rightarrow 0} \frac{ \Braket{\phi^{(2)} | \phi^{(2)}} }{h_{1,5}-h_{1,3}-2} = - \dfrac{35}{24}  .
\end{equation}

\subsection{General structure of the staggered Virasoro modules}
\label{generalstructure}

It should be clear that the path followed in the previous subsection 
can be extended to all our models. In this paper, we focus on the dense 
and dilute versions of the minimal logarithmic models $\mathcal{LM}(1,p)$ 
and $\mathcal{LM}(p,p+1)$, $p\in \mathbb{N}$. The former choice corresponds 
to $x=1/p$ and the latter to $x=p$, with $q=\mathrm{e}^{i \pi /(x+1)}$. The 
structure of the standard modules in all cases can be inferred from characters 
identities, and the staggered modules can then be built using the standards as elementary 
bricks. Using character identities, it is possible to convince oneself that the 
two following statements should hold 
\begin{itemize}
\item the  $\mathcal{LM}(1,p)$ theories ($x=1/p$) have the same 
structure as their `dual' $\mathcal{LM}(p,p+1)$ theories with $x=p$,
\item a `dilute' LCFT based on the $O(n)$ model has the same 
structure as the `dense' LCFT with the same loop fugacity $n=q+q^{-1}$.
\end{itemize}
We say that two theories have the same structure when the 
$\mathcal{P}_j$ modules in both theories have the same expression in 
terms of the simple modules $R_j$, that is to say that they are characterized by the same values of $j_1$ and $j_2$ in eq.~\eqref{eq_proj_j}.
This does not mean that the modules are the same, in particular, they may be characterized by different
indecomposability parameters, and the simple modules $R_j$ are completely different as they are over different algebras.
For example, the staggered modules in the theory with $x=1/3$ have the same structure in terms of simple modules as those of the Ising model,
given by eq.~\eqref{eq_proj_ising}.

Therefore, everything boils down to the study of the staggered modules\footnote{Note that what we call staggered modules here are nothing but the `scaling limit' of the Temperley-Lieb projective modules that arise in lattice models. They may or may not be indecomposable diamonds depending on the value of the `spin' $j$.}
of $\mathcal{LM}(p,p+1)$ theories with $x=p$. 
For such theories, we find that the staggered modules have the following form 
\begin{equation}
{\cal P}_{j}=~~~~~\left\{\begin{array}{cl}
\begin{array}{ccccc}
     &&\hskip-.7cm R_j&&\\
     &\hskip-.2cm\swarrow&\searrow&\\
     R_{j-1}&&&\hskip-.3cm R_{j+p}\\
     &\hskip-.2cm\searrow&\swarrow&\\
     &&\hskip-.7cm R_j&&
     \end{array}&\hbox{$j\equiv0$ \ (mod $\frac{p+1}{2}$) and $j>0$,}\nonumber\\
     &\nonumber\\
\begin{array}{ccccc}
     &&\hskip-.7cm R_j&&\\
     &\hskip-.2cm\swarrow&\searrow&\\
     R_{j-2}&&&\hskip-.3cm R_{j+p-1}\\
     &\hskip-.2cm\searrow&\swarrow&\\
     &&\hskip-.7cm R_j&&
     \end{array}&\hbox{$j\equiv\frac{1}{2}$ \ (mod $\frac{p+1}{2}$) and $j>1$,}\nonumber\\
     &\nonumber\\
\vdots \\
\begin{array}{ccccc}
     &&\hskip-.7cm R_j&&\\
     &\hskip-.2cm\swarrow&\searrow&\\
     R_{j-p}&&&\hskip-.3cm R_{j+1}\\
     &\hskip-.2cm\searrow&\swarrow&\\
     &&\hskip-.7cm R_j&&
     \end{array}&\hbox{$j\equiv\frac{p-1}{2}$ \ (mod $\frac{p+1}{2}$) and $j>p-1$,}\nonumber\\
     &\nonumber\\
R_j &\hbox{$j\equiv\frac{p}{2}$ \ (mod $\frac{p+1}{2}$).}\nonumber\\
     &\nonumber\\
\end{array}\right.
\end{equation}
Once again, each diamond module corresponds to a Jordan cell for $L_0$ involving the null-field $\phi_j(z)$, with 
logarithmic coupling $\beta_{1,1+2j}$ ({\it resp.} $\beta_{1+2j,1}$) in the 
dense ({\it resp.} dilute) case given by eq.~\eqref{b_formula}. For $j<p-1$, 
only the top standard module of the diamond remains in the staggered.


\section{Numerical measure of indecomposability parameters from lattice models}

\label{numericalmethod}

While the analysis of symmetries of the lattice models provides information about the general structure of  the Virasoro indecomposable modules,
getting more detailed information about the action of the Virasoro generators in these modules---such as the  numerical values of the  
indecomposability parameters---is more challenging.

There are many difficulties to overcome in this kind of analysis. One of the most important ones concerns 
the proper normalization of the state $\phi$, which obeys  $\Braket{\phi | \phi} = 0$. 
For $c=0$ and the Jordan cell of the energy-momentum tensor, Ref.~\cite{DJS}  used a trick 
that led to the determination of  
$b=-5/8$ for percolation and  $b=5/6$ for self-avoiding random walks (dilute polymers). 
We follow here a different route, and propose a more general method which allows us to study
other Jordan cells with an $A$ operator more complicated than $L_{-2}$.
The method turns out te be quite accurate so that indecomposability parameters can be
determined numerically, with almost the same precision as for critical exponents.



\subsection{Jordan cells and lattice scalar product}

\subsubsection{Spectrum and Jordan cells}

We consider a generic Jordan cell at level $h$ in $L_0$. As before, we normalize 
our states such that in the basis ($\phi$,$\psi = A \xi$), $L_0$ reads
\begin{equation}
L_0 = 
\left( \begin{array}{cc}
h & 1 \\
0 & h \end{array} \right).
\end{equation}
It is well known that the operator $L_0$ can be related to the scaling limit 
of a Hamiltonian at a critical point. For a system of length $L=2N$, we have
\begin{equation}
\displaystyle H = E_S + E^{\infty}_0 L + \frac{\pi v_F}{L} \left( L_0 - \frac{c}{24}\right) + \mathcal{O}(L^{-2}),
\end{equation}
where $E_S$ is a (non-universal) surface energy and $E_0^{\infty}$ is the bulk 
energy density. Hence, we expect non-diagonalizability also in our lattice 
Hamiltonians (or transfer matrices) that mimic the continuum limit behavior. 
The central charge and the critical exponents of a given model can be readily 
measured numerically using finite-size corrections in $L=2N$ to the 
eigenvalues of $H$. If we note $E_0$ the energy of the fundamental and $E_{\phi}$ that 
of a given excitation, we have the well-known relations \cite{CardyScaling}
\begin{subequations}
\begin{eqnarray}
\displaystyle E_0(L) &=& E_0(\infty)L + E_0^S -\frac{v_F \pi c}{24 L} + \dots \\
\displaystyle E_{\phi}(L) - E_{0}(L) &=& E_{\phi}^S- E_0^S +\frac{v_F \pi}{L} h_{\phi} + \dots
\end{eqnarray}
\end{subequations}
with most of the times $E_{\phi}^S= E_0^S$. We here assumed that all the 
conformal dimensions $h_{\phi}$ were positive; if that is not the case, 
one introduces the usual concept of effective central charge $c_{\mathrm{eff}}=c-24 h_{\rm min}$. 
Using these formulae, it is then a simple matter to identify the different 
eigenstates of the Hamiltonian which corresponds to $ \Ket{\psi}$,  $\Ket{\phi}$ 
or $\Ket{\xi}$ in the continuum limit. To do so, we use the Arnoldi algorithm \cite{Arnoldi} to 
get the eigenvalues and the corresponding Schur vectors for the lowest excitations 
of the spectrum. We then apply some variant of the Gauss-Jordan algorithm to put 
the reduced Schur (upper triangular) matrix into Jordan canonical form\footnote{This step involves choosing 
a rule to declare that two numbers $\lambda_1$ and $\lambda_2$ are equal if and only if $|\lambda_1-\lambda_2|<\varepsilon$
with typically $\varepsilon \sim 10^{-8}$.}.

Let us suppose that we identify a Jordan cell in the Hamiltonian which 
corresponds to an energy $E(L) $.
We normalize the states to prepare the comparison with CFT. In the 
basis $\lbrace \Ket{\phi^{(L)}}, \Ket{\psi^{(L)}} \rbrace$, the Hamiltonian 
for a system with $L=2N$ sites reads 
\begin{equation}
\displaystyle H^{(L)} - E_0(L) \mathrm{Id}= \frac{\pi v_F}{L} \left( \begin{array}{cc} h^{(L)} & 1  \\ 0 & h^{(L)}  \end{array} \right),
\label{eq_latticeH}
\end{equation}
where $v_F$ is the Fermi velocity and $h^{(L)} = \frac{L}{\pi v_F} (E(L) - E_0(L)) $. 
Note that we have $\lim_{L \to \infty} h^{(L)} = h$, so the matrix expression in eq.~
\eqref{eq_latticeH} goes to $L_{0}$ in the continuum limit. 

\subsubsection{Lattice scalar products}

In order to measure $\beta = \Braket{\psi | \phi}$, we first need to define a 
`scalar product' that goes to the Virasoro bilinear form in the scaling limit.
The construction of lattice `scalar products' which go to the Virasoro form 
in the continuum has already been studied in great details in Ref. \cite{DJS}. 
All scalar products must be chosen such that $L_0^{\dag} = L_0$ in the 
underlying CFT. This means that we want the Hamiltonian $H$ to be hermitian 
for these scalar products. Of course, as we deal with non-unitary theories, 
there may be negative-norm states so what we call here scalar product is nothing 
but a sesquilinear form. Obviously, the non-hermitianity of $H$ for the usual scalar 
product was the reason for its non-diagonalizability in the first place. Let us recall 
the expression of the scalar product for the different representations of the (dense/dilute) 
Temperley-Lieb algebra:
\paragraph{XXZ:}  The scalar product is the usual bilinear form on $\mathbb{C}$ without 
complex conjugation, that is, treating $q$ as a formal parameter. For example, on $L=4$ sites, 
the vector $\Ket{\phi} = \Ket{\uparrow \uparrow \downarrow \downarrow} + q \Ket{\uparrow \uparrow \uparrow \uparrow} $ 
has norm  $\Braket{\phi|\phi} = 1+q^2$. Note that if we had considered the usual scalar product 
on  $\mathbb{C}$, we would have found $1+|q|^2$ instead.
\paragraph{LOOP:} The correct scalar product is obtained gluing the mirror image of the first 
state on top of the second one. Each closed loop carries a weight $n=q+q^{-1}$. This is of 
course the usual form used in Temperley-Lieb representation theory. For instance, the scalar 
product between the two states
$\Ket{\alpha} = \Ket{
\psset{xunit=2mm,yunit=2mm}
\begin{pspicture}(0,0)(3,1)
 \psellipticarc[linecolor=black,linewidth=1.0pt]{-}(1.5,1.0)(1.5,1.42){180}{360}
 \psellipticarc[linecolor=black,linewidth=1.0pt]{-}(1.5,1.0)(0.5,0.71){180}{360}
\end{pspicture}
}$
and
$\Ket{\beta} = \Ket{
\psset{xunit=2mm,yunit=2mm}
\begin{pspicture}(0,0)(3,1)
 \psellipticarc[linecolor=black,linewidth=1.0pt]{-}(0.5,1.0)(0.5,0.71){180}{360}
 \psellipticarc[linecolor=black,linewidth=1.0pt]{-}(2.5,1.0)(0.5,0.71){180}{360}
\end{pspicture}
}$
is
$\Braket{\alpha|\beta} = \Braket{
\psset{xunit=2mm,yunit=2mm}
\begin{pspicture}(0,0)(3,1)
 \psellipticarc[linecolor=black,linewidth=1.0pt]{-}(1.5,0.0)(1.5,1.42){0}{180}
 \psellipticarc[linecolor=black,linewidth=1.0pt]{-}(1.5,0.0)(0.5,0.71){0}{180}
\end{pspicture}
|
\psset{xunit=2mm,yunit=2mm}
\begin{pspicture}(0,0)(3,1)
 \psellipticarc[linecolor=black,linewidth=1.0pt]{-}(0.5,1.0)(0.5,0.71){180}{360}
 \psellipticarc[linecolor=black,linewidth=1.0pt]{-}(2.5,1.0)(0.5,0.71){180}{360}
\end{pspicture}
}=
\psset{xunit=2mm,yunit=2mm}
\begin{pspicture}(0,0)(3,1)
 \psellipticarc[linecolor=black,linewidth=1.0pt]{-}(1.5,0.25)(1.5,1.42){0}{180}
 \psellipticarc[linecolor=black,linewidth=1.0pt]{-}(1.5,0.25)(0.5,0.71){0}{180}
 \psellipticarc[linecolor=black,linewidth=1.0pt]{-}(0.5,0.25)(0.5,0.71){180}{360}
 \psellipticarc[linecolor=black,linewidth=1.0pt]{-}(2.5,0.25)(0.5,0.71){180}{360}
\end{pspicture}
=n$.
The case with a non-zero number of strings $2j$ is treated in a similar fashion. 
Finally, in the case of the dilute $O(n)$ model, the scalar product between two states 
is chosen to be zero if the empty sites (marked as dots) are not the same. For example for $L=6$, $ \Braket{
\psset{xunit=2mm,yunit=2mm}
\psset{dotsize=2pt 0}
\begin{pspicture}(0,0)(5.5,1)
 \psellipticarc[linecolor=black,linewidth=1.0pt]{-}(2,1.0)(2,1.42){180}{360}
\psdot(2,1.0)
 \psellipticarc[linecolor=black,linewidth=1.0pt]{-}(2,1.0)(1,0.71){180}{360}
\psdot(5,1.0)
\end{pspicture}
|
\psset{xunit=2mm,yunit=2mm}
\begin{pspicture}(0,0)(5.5,1)
 \psellipticarc[linecolor=black,linewidth=1.0pt]{-}(0.5,1.0)(0.5,0.71){180}{360}
  \psdot(2,1.0)
 \psellipticarc[linecolor=black,linewidth=1.0pt]{-}(3.5,1.0)(0.5,0.71){180}{360}
\psdot(5,1.0)
\end{pspicture}
}
=
\Braket{
\psset{xunit=2mm,yunit=2mm}
\psset{dotsize=2pt 0}
\begin{pspicture}(0,0)(5.5,1)
 \psellipticarc[linecolor=black,linewidth=1.0pt]{-}(2,0.0)(2,1.42){0}{180}
\psdot(2,0.0)
 \psellipticarc[linecolor=black,linewidth=1.0pt]{-}(2,0.0)(1,0.71){0}{180}
\psdot(5,0.0)
\end{pspicture}
|
\psset{xunit=2mm,yunit=2mm}
\begin{pspicture}(0,0)(5.5,1)
 \psellipticarc[linecolor=black,linewidth=1.0pt]{-}(0.5,1.0)(0.5,0.71){180}{360}
  \psdot(2,1.0)
 \psellipticarc[linecolor=black,linewidth=1.0pt]{-}(3.5,1.0)(0.5,0.71){180}{360}
\psdot(5,1.0)
\end{pspicture}
}=n$,
whereas
$
\Braket{
\psset{xunit=2mm,yunit=2mm}
\psset{dotsize=2pt 0}
\begin{pspicture}(0,0)(5.5,1)
 \psellipticarc[linecolor=black,linewidth=1.0pt]{-}(1.5,1.0)(1.5,1.42){180}{360}
 \psellipticarc[linecolor=black,linewidth=1.0pt]{-}(1.5,1.0)(0.5,0.71){180}{360}
\psdot(4,1.0)
\psdot(5,1.0)
\end{pspicture}
|
\psset{xunit=2mm,yunit=2mm}
\begin{pspicture}(0,0)(5.5,1)
 \psellipticarc[linecolor=black,linewidth=1.0pt]{-}(0.5,1.0)(0.5,0.71){180}{360}
  \psdot(2,1.0)
 \psellipticarc[linecolor=black,linewidth=1.0pt]{-}(3.5,1.0)(0.5,0.71){180}{360}
\psdot(5,1.0)
\end{pspicture}
}=0$.

\paragraph{SUSY:} We use the usual scalar product in Fock space. There are negative 
norm states because of the use of the dual representation. For example, let us consider 
the $\mathfrak{sl}(2|1)$ case still on $L=4$ sites. A precise definition of this chain 
will be given in the following. The important point here is that each site must be occupied 
by one particle which can be either a fermion $\{ f_i, f_j^\dag \} = (-1)^{i+1} \delta_{ij}$, 
or a Schwinger boson $[ b_{i,\sigma}, b^\dag_{j,\sigma'} ] = \delta_{ij} \delta_{\sigma\sigma'}$, 
with $\sigma \in \{ \uparrow, \downarrow \}$. Let us consider the state  $\Ket{\phi} = b^\dag_{1 \uparrow} f^\dag_{2} b^\dag_{3 \downarrow} b^\dag_{4 \uparrow}\Ket{0}$. Its norm is $\Braket{\phi|\phi}= \Braket{0 \right| b_{4 \uparrow} b_{3 \downarrow} f_{2} b_{1 \uparrow} b^\dag_{1 \uparrow} f^\dag_{2} b^\dag_{3 \downarrow} b^\dag_{4 \uparrow} \left|0} = -1$ because of the 
fermionic operator $f^\dag_{2}$ of the dual representation which satisfies $\{f_{2},f^\dag_{2} \}=-1$.

\subsection{Virasoro algebra regularization on the lattice}

There is a last difficulty that one must tackle in order to define  a proper version 
of $\beta$ on the lattice. Remark that the Jordan cell in~\eqref{eq_latticeH} is invariant 
under a global rescaling of the basis states $\Ket{\phi^{(L)}} \rightarrow \alpha \Ket{\phi^{(L)}} $ 
and  $\Ket{\psi^{(L)}} \rightarrow \alpha \Ket{\psi^{(L)}} $. Unfortunately,  the scalar product 
between the states is not invariant under such transformation $ \Braket{\psi^{(L)}|\phi^{(L)}} \rightarrow |\alpha|^2 \Braket{\psi^{(L)}|\phi^{(L)}}$. Hence, we need to normalize the state $\Ket{\phi^{(L)}} $ so that it goes 
precisely to $\Ket{\phi}= A \Ket{\xi}$ in the continuum limit.

Let $\Ket{\phi^{(L)}} = \alpha \Ket{\tilde{\phi}^{(L)}}$ and 
$\Ket{\psi^{(L)}} = \alpha \Ket{\tilde{\psi}^{(L)}}$ where $\Ket{\tilde{\phi}^{(L)}}$ 
goes to $\Ket{\phi}$ when $L \rightarrow \infty$. If we knew $\alpha$, we would be 
able to compute $\Braket{\tilde{\psi}^{(L)}|\tilde{\phi}^{(L)}} \rightarrow \beta$ 
as $L \rightarrow \infty$. Note that $\Ket{\phi}$ is a null state so that we cannot 
simply normalize it, we thus need to find another way to get rid of this global 
factor $\alpha$. This is achieved using a regularization of the Virasoro generators 
on the lattice. Following Ref. \cite{KooSaleur}, for a general critical Hamiltonian 
\begin{equation}
H = - \sum_{i=1}^{L-1} h_{i}
\end{equation}
with Fermi velocity $v_F$, we define a lattice version of the $L_n$'s 
\begin{equation}
\displaystyle L^{(2N)}_n = \frac{L}{\pi} \left[ - \frac{1}{v_F} \sum_{i=1}^{L-1} (h_i - h_{\infty}) \cos \left( \frac{n i \pi}{L} \right) +  \frac{1}{v_F^2}  \sum_{i=1}^{L-2} \left[ h_i, h_{i+1} \right] \sin \left( \frac{n i \pi}{L} \right)\right] + \frac{c}{24} \delta_{n,0},
\label{eq_KooSaleur}
\end{equation}
where $h_{\infty}$ is the ground state expectation value of $h_i$. It is possible 
to show that such expression provides a good way to define the Virasoro algebra 
on the lattice in the case where $h_i=e_i$ is a Temperley-Lieb generator. In 
particular, it is possible to measure the central charge or scalar products 
of the continuum limit through the computation of Virasoro commutators. We 
shall admit that this equation remains correct even if $h_i$ is a generator 
of the dilute Temperley-Lieb algebra.
Note that these generators do not exactly satisfy the Virasoro algebra in 
the continuum limit because of anomalies: the commutators of the scaling limit do not coincide in general
with  the scaling limit of the commutators due to extra couplings to `non-scaling states' \cite{KooSaleur}. 
While the problem can be solved using a double limit procedure, in practice,
the anomalies induce extremely small errors to the lattice measurements. For 
practical purposes, formula~(\ref{eq_KooSaleur}) can thus be  used naively, 
even when commutators of multiple actions of Virasoro generators are involved.

Using  formula (\ref{eq_KooSaleur}, we are thus  able to construct a lattice version $A^{(L)}$ 
of the A operator that links $\phi$ and $\xi$. Moreover, the state $\Ket{\xi^{(L)}}$ 
is readily identified in the spectrum. We normalize it such that 
$\Braket{\xi^{(L)}|\xi^{(L)}}=1$. If we assume that 
$\Ket{\tilde{\phi}^{(L)}} =  A^{(L)} \Ket{\xi^{(L)}}$ is a correct 
lattice version of $\Ket{\phi}$, we are now ready to compute $\beta$. 
Gathering all the pieces,
we define a lattice version of $\beta$ which does not depend on $\alpha$
\begin{equation}
\displaystyle \beta^{(L)} = \dfrac{\left| \Braket{\psi^{(L)}|A^{(L)} \xi^{(L)}} \right|^2 }{\Braket{\psi^{(L)}|\phi^{(L)}}}.
\label{eq_b_lattice}
\end{equation}
When $L \rightarrow \infty$, we check that 
\begin{equation}
\displaystyle \lim_{L \rightarrow \infty} \beta^{(L)} = \lim_{L \rightarrow \infty} \dfrac{ |\alpha|^2 \left| \Braket{\tilde{\psi}^{(L)}|A^{(L)} \xi^{(L)}} \right|^2 }{|\alpha|^2 \Braket{\tilde{\psi}^{(L)}|\tilde{\phi}^{(L)}}} = \lim_{L \rightarrow \infty} \Braket{\tilde{\psi}^{(L)}|\tilde{\phi}^{(L)}}=\Braket{\psi|\phi}= \beta.
\end{equation}
We summarize our method to measure $\beta$ by the following steps:
\begin{enumerate}
\item Using exact diagonalization methods, find a Jordan basis for the first few excitations of $H$ on $L=2N$ sites.
\item Identify a Jordan cell in the spectrum of $H$ and normalize the states like in eq.~\eqref{eq_latticeH}.
\item Also identify the state $\Ket{\xi^{(L)}}$ and normalize it such that $\Braket{\xi^{(L)}|\xi^{(L)}}=1$ for the lattice scalar product.
\item Using Virasoro generators on the lattice~\eqref{eq_KooSaleur}, construct the operator $A^{(L)}$.
\item Compute $\beta^{(L)}$ using eq.~\eqref{eq_b_lattice}.
\end{enumerate}
The value of the indecomposability parameter $\beta = \lim_{L \rightarrow \infty} \beta^{(L)}$ is then computed using an extrapolation $\beta^{(L)} = \beta + A/L + B/L^2 + \dots$  We find numerically that $\beta^{(L)}$ does not depend on the chosen Temperley-Lieb representation (Loop, XXZ, or SUSY). However, it does depend on how
we realize the operator $A$ on the lattice\footnote{In general, there are infinitely many ways to realize the operator A on the lattice. In the case of $A=L_{-2}$
acting on the vacuum for example, one can use the trousers trick \cite{DJS}, or the Koo-Saleur formula~(\ref{eq_KooSaleur}) directly $A=L^{(2N)}_{-2}$, or even $A=L^{(2N)}_{-2}+ \alpha L^{(2N)}_{2}$ where $\alpha \in \mathbb{C}$. The values of $\beta^{(L)}$ computed using these different lattice realizations are different, although we expect them to yield the same result in the limit $N \rightarrow \infty$.}, this is why we were able to improve the results of Ref. \cite{DJS}.


\section{Numerical results}

We present in this section our numerical results for indecomposability parameters in different examples of LCFT.

\subsection{The case $x=1$ ($c=-2$): Dense Polymers}

We begin with the case of the XX(Z) chain of even length $2N$, 
with $q=i$ so $c=-2$. The expansion of the partition function 
in terms of dense loops describes dense polymers as the weight 
for a closed loop is $n=0$. This theory is relevant for the 
description of spanning trees \cite{spanningtrees} (up to a duality transformation).
 It is also   related with the description of abelian sandpile models, 
although in the latter case different values of the indecomposability parameters have been found \cite{Sandpile2,Sandpile}.
In the SUSY language, this model also corresponds to $\mathfrak{gl}(m|m)$ 
(super)spin chains and to non-linear sigma models with target space 
$\mathbb{CP}^{m-1|m}$ at $\theta = \pi$ \cite{RS1}. For $m=1$, we 
get the $\mathfrak{gl}(1|1)$ spin chain which is a free fermion system. 
Indeed, at $m=1$ the $U_q(\mathfrak{sl}_2)$ XX spin chain and the supersymmetric 
$\mathfrak{gl}(1|1)$-invariant chain coincide and everything can be 
reformulated in terms of free-fermion generators obeying 
$\lbrace f_i,f_j \rbrace=0$ and $\lbrace f_i,f^\dagger_j \rbrace=(-1)^{i+1} \delta_{i,j}$. 
Within this representation, the Temperley-Lieb generators read
\begin{equation}
\displaystyle e_i = (f^\dagger_i + f^\dagger_{i+1})(f_i + f_{i+1}).
\end{equation}
The corresponding continuum limit is a symplectic fermions 
theory \cite{Kausch1, Kausch2} with action
\begin{equation}
\displaystyle S = \frac{1}{2 g_{\sigma}^2} \int \mathrm{d}^2 r \  \partial_{\mu} \xi^\dagger \partial^{\mu} \xi.
\end{equation}
This theory is probably one of the best understood LCFT. There 
are $4$ different fields at level $0$ which are organized into 
a diamond indecomposable module of $\mathfrak{gl}(1|1)$. All the Virasoro staggered modules 
can be constructed as fermionic excitations of these states at 
level $0$. A few indecomposability parameters have been computed 
using the Kausch-Gaberdiel algorithm \cite{KauschGaberdiel}. One can 
also readily compute the same parameters using the free fermion 
representation.

We show here how to implement our lattice approach to this model. The Kac formula at $x=1$ reads
\begin{equation}
\displaystyle h_{r,s} = \frac{ \left(2r - s \right)^2 - 1}{8},
\end{equation}
and extending the generic results to the case $q=i$, we find 
that the partition function of the $\mathfrak{gl}(1|1)$ spin chain is
\begin{equation}
 Z = \sum_{j=0}^{\infty} (2j+1) \frac{q^{(2j-1)^2/8}-q^{(2j+3)^2/8} }{\eta(q)} = \sum_{j=1}^{\infty} j (2 \chi_{j,1}+\chi_{j+1,1}+\chi_{j-1,1}).
\end{equation}
Recall that $\chi_{j,1} = \chi_{1,1+2j} $ is the irreducible 
character of the Virasoro simple module with $h=h_{j,1}=h_{1,1+2j}$.
Using the general results presented in section \ref{generalstructure} with $p=1$ 
and $j = \frac{1}{2} (\mathrm{mod} \ 1)$, 
we see that there is no indecomposability for $L$ odd ($j$ half-integer) 
so that the lattice Hamiltonian remains fully diagonalizable in this 
case. For $j \in \mathbb{N}$, the staggered Virasoro modules have the 
following subquotient structure and a basis in terms of fields
\begin{equation}
\mathcal{P}_{j}=
 \begin{array}{ccccc}
      &&\hskip-.7cm R_{j} &&\\
      &\hskip-.2cm\swarrow&\searrow&\\
      R_{j-1} &&&\hskip-.3cm R_{j+1} \\
      &\hskip-.2cm\searrow&\swarrow&\\
      &&\hskip-.7cm R_{j} &&
\end{array}
=
 \begin{array}{ccccc}
      &&\hskip-.7cm \psi^{(j)} &&\\
      &\hskip-.2cm\swarrow&\searrow&\\
      \xi^{(j)} &&&\hskip-.3cm \rho^{(j)} \\
      &\hskip-.2cm\searrow&\swarrow&\\
      &&\hskip-.7cm \phi^{(j)}&&
\end{array}.
\end{equation}
The modules $\mathcal{P}_{j}$ are completely characterized by the 
logarithmic couplings $\beta_{1,1+2j}$, and each module corresponds to a Jordan cell
\begin{align}
L_0 \phi^{(j)} & = h_{1,1+2j} \phi^{(j)} \\ \notag
L_0 \psi^{(j)} & = h_{1,1+2j} \psi^{(j)} + \phi^{(j)} \\ \notag
\phi^{(j)} & = A_j \xi^{(j)} \\ \notag
A_j^{\dagger} \psi^{(j)} & = \beta_{1,1+2j} \xi^{(j)} .
\end{align}
Note that we choose a different convention for the operator $A$ than 
eq.~\eqref{eq_Aconv} in order to match \cite{KauschGaberdiel} as we 
normalize it such that $A_j = L^{j-1}_{-1}+\dots$ Using this convention 
and eq.~\eqref{b_formula}, we were able to conjecture a general formula 
for the indecomposability parameter
\begin{equation}
\displaystyle \beta_{1,1 + 2j} = - \lim_{x \rightarrow 1} \frac{ \Braket{\xi |A^\dag A |\xi}}{h_{1,1+2j}-h_{1,1+2(j-1)}-(j-1)}= - \frac{[(2j-3)!]^2}{4^{j-2}}(j-1), \ \ j \in \mathbb{N}\setminus\{0,1\}. 
\end{equation}
We checked this equation up to $j=16$ using eq.~\eqref{b_formula}. 
Note also that in other cases can similar explicit formulae be obtained from
 eq.~\eqref{b_formula}; we shall report on this point and compare
our results with the literature (see {\it e.g.} \cite{MathieuRidout1,Kytola}) in a separate publication \cite{AVJS}.

We can try to measure these numbers numerically using the method described 
in section \ref{numericalmethod}. To do so, we need to construct the lattice 
version of the $A_j$ operator using eq.~\eqref{eq_KooSaleur} with 
$h_i=e_i$, $h_{\infty}= 2/ \pi$ and $v_F=2$. For $j=2,3,4$, these operators read
\begin{subequations}
\begin{eqnarray}
      A_{2} &=& L_{-1} , \ \ \beta_{1,5}=-1 \\
      A_{3} &=& L^2_{-1} - 2 L_{-2}, \ \  \beta_{1,7}=-18 \\
      A_{4} &=& L^3_{-1} - 8 L_{-2} L_{-1} +12 L_{-3}, \ \  \beta_{1,9}=-2700
\end{eqnarray}
\end{subequations}
We measured these three parameters in the spin sectors $S_z=1, 2$ and $3$, 
respectively. The results are presented in Tab. \ref{tab_b_dense_polymers}, 
in very good agreement with the theoretical expectation. We computed these 
numbers using both supersymmetric (XX) and geometrical representations of 
the Temperley-Lieb algebra and obtained the same values in finite size.

\begin{table}
\begin{center}
\begin{tabular}{|c|c|c|c|}
  \hline
  $L = 2 N $ & $\beta_{1,5}$ & $\beta_{1,7}$ & $\beta_{1,9}$\\
  \hline
  8 & -0.937759 & -13.3574 &\\
  10 & -0.959708 & -14.8908 & -1518.37 \\
  12 & -0.971844 & -15.7936 & -1805.43 \\
  14 & -0.979236 & -16.3612 & -2013.66 \\
  16 & -0.984064 & -16.7384 & -2157.86 \\
  18 & -0.987388 & -17.0006 & -2262.59 \\ 
  20 & -0.989771 & -17.1898 & -2340.51 \\ 
  22 & -0.991539 & -17.3304 & -2399.80 \\
  24 &  &  & -2445.81 \\
  \hline
  $\infty$ & -1.0000 $\pm$ 0.0002 & -18.0(0) $\pm$ 0.05 & -27(00) $\pm$ 25\\
  \hline
  Exact & $-1$ & $-18$ & $-2700$\\
  \hline

\end{tabular}
\end{center}
\caption{Measure of indecomposability parameters in the XXZ spin chain at $q=i$.}
  \label{tab_b_dense_polymers}
\end{table}

\subsection{The cases $x=2$ ($c=0$) and $x=1/2$ ($c=-7$): Percolation and $\mathfrak{sl}(2|1)$ superspin chain}
We now deal with a slightly more complicated example, the antiferromagnetic 
$\mathfrak{sl}(2|1)$ superspin chain. This chain is known to be equivalent 
to the classical percolation problem, and arises naturally in the context 
of the Spin Quantum Hall Effect \cite{SQHE}. The percolation problem has 
$c=0$ and can also be formulated in terms of geometrical clusters or the XXZ 
spin chain at $q=\mathrm{e}^{2i \pi/3}$. We also consider the ferromagnetic 
version, with central charge $c=-7$.

\subsubsection{$\mathfrak{sl}(2|1)$-invariant spin chain}
We consider a chain of alternating fundamental and dual representations 
of the Lie superalgebra $\mathfrak{sl}(2|1)$ with `Hilbert' space 
$\mathcal{H} = (\square \otimes \bar{\square})^{\otimes N}$. For more 
details about $\mathfrak{sl}(2|1)$ and its representation theory, we 
refer the interested reader to the literature (see {\it e.g.} Ref. \cite{sl21}). 
The Hilbert space on one specific site is spanned by three independent states 
so the whole Hilbert space has dimension $3^{L}=9^{N}$. On every site, we introduce 
two boson operators $[ b_{i,\sigma}, b^\dag_{j,\sigma'} ] = \delta_{ij} \delta_{\sigma\sigma'}$, 
where $\sigma \in \{ \uparrow, \downarrow \}$, and one fermion 
$\{ f_i, f_j^\dag \} = (-1)^{i+1} \delta_{ij}$. We add a constraint 
to the system so that there cannot be more than one particle by site. 
The representation on the site $i$ is $\mathbb C^3  \simeq \mathrm{Span} \{ f_i^\dag \Ket{0}, b_{i,\uparrow}^\dag \Ket{0}, b_{i,\downarrow}^\dag \Ket{0} \}$ and corresponds to the 
fundamental $\square$ for $i$ odd and to the dual $\bar{\square}$ 
otherwise. The invariant coupling is chosen to be the Casimir in the 
tensor product representation of the sites $i$ and $i+1$.
\begin{equation}
\displaystyle e_i = (b_{i+1,\downarrow}^\dag b_{i,\uparrow}^\dag + b_{i+1,\uparrow}^\dag b_{i,\downarrow}^\dag + (-1)^{i+1} f_{i+1}^\dag f_i^\dag) (b_{i,\uparrow} b_{i+1,\downarrow} + b_{i,\downarrow} b_{i+1,\uparrow} + (-1)^{i+1} f_i f_{i+1}).
\end{equation}
It furnishes a representation of the Temperley-Lieb algebra 
with $n=1$. The Hamiltonian reads
\begin{equation}
\displaystyle H= \pm \sum_{i=1}^{L-1} e_i,
\end{equation}
where the minus sign corresponds to percolation ($c=0$) and 
the plus sign to the ferromagnetic case ($c=-7$). There are 
two good quantum numbers $S_z$ and $B$ that we can use to label the states 
\begin{subequations}
\begin{eqnarray}
S_z &=& \frac{1}{2}\sum_{i=1}^{L} (b_{i,\uparrow}^\dag b_{i,\uparrow} - b_{i,\downarrow}^\dag b_{i,\downarrow}), \\
B &=& \sum_{i=1}^{L} \left( (-1)^{i+1} \frac{b_{i,\uparrow}^\dag b_{i,\uparrow} + b_{i,\downarrow}^\dag b_{i,\downarrow}}{2} + f^\dag_i f_i \right).
\end{eqnarray}
\end{subequations}
Finally, note that the low energy physics of this chain can 
be described by a non-linear sigma model with target space $\mathbb{CP}^{1|1}$ at $\theta = \pi$ \cite{RS1}. 

\begin{table}
\begin{center}
\begin{tabular}{|c|c|c|c|c|c|}
\cline{1-6}
  \multicolumn{4}{|c|}{Percolation ($q=\mathrm{e}^{i \pi/3}$)} & \multicolumn{2}{|c|}{Ferro $\mathfrak{sl}(2|1)$ ($q=\mathrm{e}^{2i \pi/3}$)} \\ \cline{1-6}
  $L$ & $\beta_{1,4}$ & $ L$ & $\beta_{1,5}$ & $L$ & $\beta_{1,7}$\\
  \hline
  7 & -0.471874  & 8 & -0.609088 & 8 &  -1.57616\\
  9 & -0.476386  & 10 & -0.605858 & 10 &  -1.77278\\
  11 & -0.479983  & 12 & -0.606403 & 12  & -1.87138  \\
  13 & -0.482724  & 14 & -0.607775 & 14  & -1.92567  \\
  15 & -0.484837 & 16  & -0.609226 & 16  & -1.95770 \\ 
  17 & -0.486503 & 18  & -0.610561 & 18  & -1.97762\\ 
  19 & -0.487845 & 20 & -0.611738  & 20 & -1.99049\\
  21 & -0.488946 & 22 & -0.612764  & 22 & -1.99906\\
  \hline
  $\infty$ & -0.5000 $\pm$  0.0001 &$\infty$ &   -0.6249 $\pm$ 0.0005  & $\infty$ & -2.00 $\pm$  0.005  \\
  \hline
  \hline
  Exact & -1/2 &Exact & -5/8 = 0.625 & Exact & $-2$\\
  \hline
\end{tabular}
\end{center}
\caption{Measure of indecomposability parameters in $c=0$ and $c=-7$ theories.}
  \label{tab_perco}
\end{table}

\subsubsection{Measure of indecomposability parameters}
First of all, let us focus on the first known indecomposability 
parameter which concerns the stress energy tensor 
in the percolation problem. 
The Kac formula with $x=2$ reads
\begin{equation}
\displaystyle h_{r,s} = \frac{ \left(3r - 2s \right)^2 - 1}{24},
\end{equation}
and the values appearing in the spectrum are
\begin{equation}
\displaystyle h_{1,1+2j} = \frac{j(2j-1)}{3}.
\end{equation}
The partition function of the $q=\mathrm{e}^{i \pi/3}$ XXZ 
spin chain with an even number of sites reads
\begin{equation}
\displaystyle Z = \sum_{j=0}^{\infty} (2j+1) \frac{q^{(4j-1)^2/24}-q^{(4j+5)^2/24} }{\eta(q)}.
\end{equation}
For the $\mathfrak{sl}(2|1)$-invariant chain, one 
replaces $2j+1$ by $[2j+1]_{q'}$\footnote{We define the q-analog as $[n]_q = \frac{q^{n}-q^{-n}}{q-q^{-1}}$.} with $q'+q'^{-1}=3$ \cite{RS3,RS2}.
As we already argued in details in section \ref{subsec_catastrophe}, 
the stress energy tensor at $c=0$ must have a logarithmic partner $t(z)$ 
which corresponds to the field $\psi(z)$ with our notations.
It is now well admitted that the indecomposability parameter in this 
case is $\beta_{1,5}=-5/8$. This number was also measured numerically 
in Ref. \cite{DJS}. The Jordan cells equations in that case read
\begin{align}
L_0 \Ket{T} & = 2 \Ket{T} \\ \notag
L_0 \Ket{t} & = 2 \Ket{t} + \Ket{T} \\ \notag
\Ket{T} & = L_{-2} \Ket{0} \\ \notag
L_2 \Ket{t} & = b \Ket{0} .
\end{align}
Using the general structure (section \ref{generalstructure}), we see that 
these states are organized into the following diamond structure
\begin{equation}
\mathcal{P}_2=
 \begin{array}{ccccc}
      &&\hskip-.7cm R_2&&\\
      &\hskip-.2cm\swarrow&\searrow&\\
      R_0&&&\hskip-.3cm R_3 \\
      &\hskip-.2cm\searrow&\swarrow&\\
      &&\hskip-.7cm R_2&&
\end{array}
=
 \begin{array}{ccccc}
      &&\hskip-.7cm t&&\\
      &\hskip-.2cm\swarrow&\searrow&\\
      I&&&\hskip-.3cm \rho \\
      &\hskip-.2cm\searrow&\swarrow&\\
      &&\hskip-.7cm T&&
\end{array}
\end{equation}
We measured $\beta_{1,5}$ for various Temperley-Lieb representations. In the 
XXZ spin chain at $q=\mathrm{e}^{i\pi/3}$, the Jordan cell occurs 
in the sector $S_z=0$, while in the $\mathfrak{sl}(2|1)$ SUSY case, 
the Jordan cell for $T$ is to be found in the sector $(S_z,B)=(0,0)$. 
We recall that the lattice indecomposability parameter given by eq.~\eqref{eq_b_lattice} does not depend on the chosen representation. 
In the geometrical setup of percolation as dense loop gas with 
fugacity $n=1$, the Hamiltonian remains fully diagonalizable and 
there is no coefficient to measure here. Nevertheless, it is still 
possible to slightly deform it \cite{DJS} so that Jordan cells appear, 
and in this case we find the same values as in the other representations. 
The results (Tab. \ref{tab_perco}) are in excellent agreement with 
the prediction $b = -5/8$ and significantly improve the precision of the results 
obtained from the trousers trick \cite{DJS}. 

One can also measure the indecomposability parameters 
$\beta_{1,1+2j}$ with $j$ half-integer from odd-length chains. 
For instance, for $L$ odd, there is a Jordan cell at level 1 
that corresponds to the Virasoro staggered module $\mathcal{P}_\frac{3}{2}$.  
This Jordan cell occurs in the sector $S_z=1/2$, we call as usual 
$\xi$ the unique state with $h=0$ in this sector, and $\psi$ and 
$\phi$ the states with $h=1$. We normalize $\xi$ such 
that $\Braket{\xi|\xi}=1$. The OPE formula~\eqref{b_formula} 
gives a logarithmic coupling $\beta_{1,4} = \Braket{\phi | \psi} = -1/2$; 
this value was also found by Mathieu and Ridout using different 
methods \cite{MathieuRidout,MathieuRidout1}. We can measure this 
coefficient on the lattice using the same method; once again, 
the results are in excellent agreement with the theoretical 
expectation (Tab. \ref{tab_perco}).

Another case of interest is $q=\mathrm{e}^{2 i \pi/3}$, with a 
central charge $c=-7$. This model corresponds to an $\mathfrak{sl}(2|1)$ 
spin chain with ferromagnetic couplings. Graphical expansion of 
the partition function yields a loop model with fugacity $n=-1$. 
The conformal dimensions appearing in the spectrum are given 
by the Kac formula for $x=1/2$
\begin{equation}
\displaystyle h_{1,1+2j} = \frac{j(j-2)}{3}.
\end{equation}
The first interesting Jordan cell arises at level 1, and 
corresponds to the staggered module
\begin{equation}
\mathcal{P}_3=
 \begin{array}{ccccc}
      &&\hskip-.7cm R_3&&\\
      &\hskip-.2cm\swarrow&\searrow&\\
      R_2&&&\hskip-.3cm R_5 \\
      &\hskip-.2cm\searrow&\swarrow&\\
      &&\hskip-.7cm R_3&&
\end{array}.
\end{equation}
In this case, eq.~\eqref{b_formula} with $A=L_{-1}$ yields 
$\beta_{1,7} = -2$. This coefficient was also computed by Kausch and Gaberdiel \cite{KauschGaberdiel}
thanks to the Nahm-Gaberdiel-Kausch algorithm.
This cell occurs in the sectors $S_z=-2,-1,1,2$ 
of the XXZ spin chain. Measures of $\beta_{1,7}$ in all these sectors 
yield values in good agreement\footnote{Extrapolation are done 
fitting data by $\beta+ A/L + B/L^2$, so the resulting fitting curves 
need not be monotonic; this is particularly obvious in this case.} 
with $\beta_{1,7} = -2$ (see Tab. \ref{tab_perco}). As in the 
other cases, the lattice values of $\beta_{1,7}$ do not 
depend on the chosen representation.

\begin{table}
\begin{center}
\begin{tabular}{|c|c|}
  \hline
  $L = 2 N $ & $\beta_{1,5}$ \\
  \hline
  8 & -1.26986  \\
  10 & -1.29548  \\
  12 & -1.31743 \\
  14 & -1.33489 \\
  16 & -1.34876 \\
  18 & -1.35993 \\
  20 & -1.36905 \\
  22 & -1.37663 \\
  \hline
  $\infty$ & -1.4582(8) $\pm$ 0.0001 \\
  \hline
  Exact & $-35/24 \simeq -1.4583$ \\
  \hline
\end{tabular}
\end{center}
\caption{Measure of $\beta_{1,5}$ in the XXZ spin chain at $q=\mathrm{e}^{i \pi/4}$.}
\label{tab_ising}
\end{table}

\subsection{The case $x=3$ ($c=\frac{1}{2}$): Logarithmic Ising model}
Let us consider the case $q=\mathrm{e}^{i \pi/4}$, which corresponds 
to a central charge $c=\frac{1}{2}$. It corresponds to a dense loop 
model with fugacity $n=\sqrt{2}$; this is of course the celebrated 
(logarithmic) Ising model $\mathcal{LM}(3,4)$. 
The spectrum is given by the Kac formula at $x=3$
\begin{equation}
\displaystyle h_{1,1+2j} = \frac{j(3j-1)}{4}.
\end{equation}
The partition function of the $U_{q=\mathrm{e}^{i \pi/4}}(\mathfrak{sl}_2)$ XXZ spin chain reads
\begin{equation}
\displaystyle Z = \sum_{j=0}^{\infty} (2j+1) \frac{q^{(1- 6 j)^2/48}-q^{(7+6j)^2/48} }{\eta (q)}.
\end{equation}
The algebraic structure of the continuum limit was studied in details 
in section \ref{subsec_ising}. We measured numerically the 
indecomposability parameter associated to the module~\eqref{eq_projisingex}. 
The results are shown in Tab.~\ref{tab_ising} and are in excellent 
agreement with the expected result $\beta_{1,5}=-35/24$ (see eq.~\eqref{eq_b_ising}).

\subsection{Another theory at $c=0$: $O(n \rightarrow 0)$ model and dilute polymers}

\begin{table}
\begin{center}
\begin{tabular}{|c|c|}
  \hline
  $L = 2 N $ & $\beta_{3,1}$ \\
  \hline
  4 & 0.021029  \\
  6 & 0.145101  \\
  8 & 0.276585 \\
  10 & 0.382046  \\
  12 & 0.463292 \\
  14 & 0.526436 \\
  \hline
  $\infty$ & 0.9 $\pm$ 0.1  \\
  \hline
  Exact & 5/6 $\simeq$ 0.8333 $\dots$ \\
  \hline
\end{tabular}
\end{center}
\caption{Measure of $\beta_{3,1}$ in the $O(n\rightarrow0)$ model. The convergence 
is quite poor compared to the Temperley-Lieb case, but the result is not 
so bad given that the precision we obtain is roughly the same as for the 
first critical exponents. The result is consistent with the known value $b=\frac{5}{6}$.}
\label{tab_0n}
\end{table}

Finally, we present here an example of measure of indecomposability 
parameters in `dilute' LCFTs. We study the $O(n \rightarrow 0)$ model, 
which is known to be relevant for the physics of dilute polymers. 
We can also formulate this model using supersymmetry, in terms of 
$\mathfrak{osp}(2m|2m)$-invariant spin chain and to non-linear sigma 
models with supersphere target space $S^{2m-1|2m}$ \cite{RS1}. 
The partition function of the $S=1$ $U_q(\mathfrak{sl}_2)$-invariant chain 
reads 
\begin{equation}
\displaystyle Z = \sum_{j=0}^{\infty} (2j+1) \frac{q^{h_{1+2j,1}}-q^{h_{-1-2j,1}} }{P (q)} = \sum_{j=0}^{\infty} (2j+1) \frac{q^{(1+6j)^2/24}-q^{(5+6j)^2/24} }{\eta (q)},
\end{equation}
where the exponents appearing in the spectrum 
are now in the first column $h_{1+2j,1}$ of the Kac table.

We are interested in the stress energy tensor $T$ in this theory, 
which is primary in this case as the central charge is $c=0$. As 
in the percolation problem, $T$ has a logarithmic partner that we 
call $t$. However, while in the percolation theory $t$ had to be 
somehow identified with the field $\Phi_{1,5}(z)$, it corresponds 
here to $\Phi_{3,1}(z)$ with thus a completely different 
indecomposability parameter $\beta_{3,1}=5/6$. We would like 
to measure this coefficient from the lattice model defined 
eq.~\eqref{hamOn}.  Unfortunately, the finite-size convergence 
of the exponents is very poor, even the central charge cannot 
be properly measured in this case. This is probably due to the 
absence of integrable K-matrices in our model; a similar phenomenon was observed 
in Ref. \cite{KooSaleur2}. Of course, one could add K-matrices at the edges 
in order to improve the convergence. Nevertheless, it is not clear to us how to adapt eq.~\eqref{eq_KooSaleur} in that case. 
Note also that the Hilbert space is much 
larger here so the accessible sizes are relatively small.

Nevertheless, we can still hope to deduce a rough estimate of 
the indecomposability parameter $b$ for the stress energy 
tensor here. Since the action of the operator $L_{2}$ on the 
vacuum gives 0, we define the operator $A^{(L)}$ as
\begin{equation}
\displaystyle L^{(L)}_{2} + L^{(L)}_{-2} = - \frac{2L}{\pi v_F} \sum_{i=1}^{L-1} (h_i-h_{\infty}) \cos \left[ \frac{2 i \pi}{L} \right].
\end{equation}
Of course, we could also have used $A^{(L)}=L^{(L)}_{-2}$ but 
as it turns out, the convergence is better with $L^{(L)}_{2} + L^{(L)}_{-2}$.
Measure of $b$ from this formula are shown in Fig.~\ref{tab_0n}. 
Although the convergence is clearly not as good as for the 
previous examples, the result is consistent with the value $5/6$.

\renewcommand{\arraystretch}{2.5}
\begin{table}
\begin{center}\small
\vspace{-2cm}
\begin{turn}{90}
\begin{tabular}{|c||c|c|c|c|c|c|c|}
  \hline
   \bf{Dense LCFTs}& $\beta_{1,3}$ & $\beta_{1,4}$ & $\beta_{1,5}$ & $\beta_{1,6}$ & $\beta_{1,7}$ & $\beta_{1,8}$ & $\beta_{1,9}$ \\
  \hline
   $\mathcal{LM}(1,5)$ ($x=\dfrac{1}{4}$)  &  $\proja$ & $\proja$  & $\projb$  & $\Diamond$ & $\Diamond$ & $\Diamond$ & $\Diamond$   \\
  \hline
   $\mathcal{LM}(1,4)$ ($x=\dfrac{1}{3}$)   & $\proja$ & $\projb$   & $\Diamond$ &   $\Diamond$ &  $\Diamond$  & $\projb$  &  $-3$ \\
  \hline
   $\mathcal{LM}(1,3)$ ($x=\dfrac{1}{2}$)  & $\projb$ &  $\Diamond$ &  $\Diamond$  & $\projb$   & $-2$ & $8$ & $\projb$   \\
  \hline
  Dense Polymers $\mathcal{LM}(1,2)$  & $\Diamond$  & $\projb$   & $-1$ & $\projb$   & $-\dfrac{9}{2}$ & $\projb$   &  $-\dfrac{75}{4}$   \\
  \hline
  Percolation $\mathcal{LM}(2,3)$   & $\projb$  & $-\dfrac{1}{2}$ & $-\dfrac{5}{8}$ & $\projb$   & $-\dfrac{35}{3}$ & $-\dfrac{13475}{216}$ & $\projb$   \\
  \hline
   Ising  $\mathcal{LM}(3,4)$  & $\proja$ & $\projb$   & $-\dfrac{35}{24}$ &  $-\dfrac{13475}{243}$ & $-\dfrac{49049}{17496}$  & $\projb$   &  $-\dfrac{40415375}{944784}$  \\
  \hline
   Tricritical Ising $\mathcal{LM}(4,5)$ & $\proja$ & $\proja$ & $\projb$  & $-\dfrac{693}{100}$ & $-\dfrac{6114399291}{1078465600}$ & $-\dfrac{91820268514045071}{253871477862400}$ & $-\dfrac{21676129054392267}{1644513366760000}$  \\
  \hline
   3-state Potts $\mathcal{LM}(5,6)$  & $\proja$ & $\proja$ & $\proja$ & $\projb$  & $-\dfrac{676039}{59895}$ & $-\dfrac{22502936626745344}{562010429701125}$ & $-\dfrac{745930435583727415172151}{24227875374038666253125}$  \\
  \hline
  \hline
   \bf{Dilute LCFTs}  & $\beta_{3,1}$ & $\beta_{4,1}$ & $\beta_{5,1}$ & $\beta_{6,1}$ & $\beta_{7,1}$ & $\beta_{8,1}$ & $\beta_{9,1}$ \\
  \hline
  Dilute Polymers $\mathcal{LM}(2,3)$  & $\dfrac{5}{6}$ & $\projb$  & $\dfrac{67375}{676}$ &  $\projb$ & $\dfrac{106462606250}{116550867}$  &$\projb$  & $\dfrac{141745038705442046875}{26751761399366832}$  \\
  \hline
  $O(n\rightarrow1)$ Ising $\mathcal{LM}(3,4)$  & $\projb$  & $\dfrac{175}{12}$ & $\dfrac{49049}{15552}$ & $\projb$   & $\dfrac{88913825}{229842}$ &   $\dfrac{362318037060948052068359375}{6798093588426728083456}$ & $\projb$   \\
  \hline
\end{tabular}

\end{turn}
\end{center}
\caption{
Indecomposability parameters of minimal chiral Logarithmic 
Conformal Field Theories. We consider two different types of 
LCFTs depending on whether the underlying loop models are in 
a dilute or dense phase. Each parameter $\beta$ corresponds 
to a diamond-shaped staggered Virasoro module. We use different 
symbols when $\beta$ is not defined: $\proja$ standard indecomposable 
module, $\projb$ irreducible (simple) module, or $\Diamond$ $L_0$-Jordan cell with no
interesting indecomposability parameter to compute.
}
\renewcommand{\arraystretch}{1.0}
\label{general_b_figure}
\end{table}

\subsection{Remarks on descendants}

We have seen that using the structure of staggered modules 
over Virasoro which arise in a given theory, one can predict 
the whole structure of Jordan cells in the continuum limit. 
However, this does not take into account descendants: there 
is a whole pyramid of Jordan cells associated with each 
(primary) Jordan cell. To be more precise, if there is a 
Jordan cell mixing two operators $\psi$ and $\phi$ with 
parameter $\beta$, then we can expect Jordan cells for all 
the descendants as is readily shown using the commutation 
relations of the $L_n$'s. The resulting indecomposability 
parameters are not independent and can be deduced from 
the knowledge of $\beta$. For instance, let us consider 
the operators $\psi^{(-n)} = L_{-n} \psi $ and 
$\phi^{(-n)} = L_{-n} \phi$ in the case $A=L_{-2}$. In the basis 
$( \psi^{(-n)},\phi^{(-n)} )$, $L_0$ reads
\renewcommand{\arraystretch}{1.0}
\begin{equation}
\displaystyle L_0 = \left( \begin{array}{cc} h+n & 1  \\ 0 & h+n  \end{array} \right),
\end{equation}
where $h$ is the conformal weight of $\psi$ and $\phi$. 
Let $\beta^{(-n)} = \Braket{\psi^{(-n)}|\phi^{(-n)}}$. Using 
the Virasoro algebra, one can show that this coupling is given by
\begin{equation}
\displaystyle \beta^{(-n)} = \left( \frac{c}{12} n(n^2-1) + \frac{c}{2} \delta_{n,2} + 4 n \right) \beta.
\end{equation}
Of course, there are similar formulae for other kinds of descendants and for other $A$ operators. 
We remark that these results are compatible with eq.~\eqref{b_formula}. 
We measured indecomposability parameters for descendants in some cases 
(results not shown here) and found a good agreement with the previous 
considerations.


\section{Conclusion}

Pushing further the analysis of \cite{RS3}, we have shown in this paper that it is possible to investigate the 
fine structure of indecomposable Virasoro modules in LCFTs using numerical analysis of 
a certain type of lattice models. 
Our method is general enough to be adapted to many cases, 
and the precision reached is almost as good as for critical exponents. 

We have restricted to the simplest type of boundary conditions for the LCFTs, but extension to more complicated cases is possible using more complicated lattice models, based for instance on the blob  algebra \cite{blob1,blob2,blob3} (or the one and two-boundary Temperley-Lieb algebra). More interestingly maybe, we believe that the method can be extended to the periodic case as well and thus should provide a powerful tool
to investigate the structure of bulk LCFTs, where very little seems to be known at present. We will report on all these questions soon.

Going back to the values of the indecomposability parameters, we also argued that they  can be
inferred from a simple heuristic argument relying on OPEs. This did not seem to be known, and suggests revisiting the bulk problem as well, by 
systematically considering  LCFTs as the limit of usual, non-logarithmic, CFTs. This will also be discussed elsewhere.

Finally, we summarize our results with a table of the first few indecomposability 
parameters for the minimal logarithmic models $\mathcal{LM}(p,p')$. 
We focus on the series $\mathcal{LM}(1,p)$ and $\mathcal{LM}(p,p+1)$, 
and we consider the two versions `dense' and `dilute' of each theory. 
Using the general structure of section \ref{generalstructure} 
and eq.~\eqref{b_formula}, one can analyze the indecomposability 
parameters in a systematic fashion \cite{AVJS}. The operators $A= L_{-n}+\dots$
are generated using the null-vector condition, and are normalized 
as in eq.~\eqref{eq_Aconv}. The results are gathered in Tab. 
\ref{general_b_figure}. 

For a given dense ({\it resp.} dilute) 
$\mathcal{LM}(p,p')$ theory, we denote $\beta_{1,1+2j}$ 
({\it resp.}  $\beta_{1+2j,1}$) the logarithmic coupling associated 
with the Jordan cell at level $h_{1,1+2j}$ ({\it resp.} $h_{1+2j,1}$). 
Of course, it may happen that this Jordan cell does not exist, or 
there may not be any interesting coefficient to measure (this is 
the case for the first few Jordan cells in  $\mathcal{LM}(1,p)$ theories), in which 
cases we use different symbols. Some of the parameters we find are 
quite complicated irreducible fractions, and the simplicity 
of the results sometimes depends on the normalization choice for $A$. 
Note that this table contains only a small fraction of the 
couplings that eq.~\eqref{b_formula} allows to compute. In principle, 
the formula~\eqref{b_formula} could be applied to obtain any indecomposability 
parameter for a given theory. The limitation obviously comes from computing 
Virasoro commutators. We shall report on all this in  \cite{AVJS}.

\vspace{10pt}

{\bf Acknowledgments.} \hspace{5pt} We are grateful to R. Bondesan, A.M. Gainutdinov,  A. Lazarescu, N. Read and  V. Schomerus  for fruitful
discussions. We also thank J. Dubail, M. Ridout and P. Ruelle for useful comments on the preprint version of this manuscript.
This work was supported by the Agence Nationale de la Recherche (grant ANR-10-BLAN-0414).

\end{document}